%% file: arxiv.tex
\documentclass{aastex6}

\shortauthors{Tran et al.}
\shorttitle{}

\begin{document}

\newcommand{\hubble}{{\it Hubble Space Telescope}}
\newcommand{\spitzer}{{\it Spitzer}}
\newcommand{\Lya}{Ly$\alpha$}
\newcommand{\Hbeta}{H$\beta$}
\newcommand{\Halpha}{H$\alpha$}
\newcommand{\Halphased}{H$\alpha_{\rm star}$}
\newcommand{\Halphagas}{H$\alpha_{\rm HII}$}
\newcommand{\Halphacut}{2~\Msunyr}
\newcommand{\NII}{\hbox{[{\rm N}\kern 0.1em{\sc ii}]}}
\newcommand{\SII}{\hbox{[{\rm S}\kern 0.1em{\sc ii}]}}
\newcommand{\OII}{\hbox{[{\rm O}\kern 0.1em{\sc ii}]}}
\newcommand{\OIII}{\hbox{[{\rm O}\kern 0.1em{\sc iii}]}}
\newcommand{\mipsmu}{$24\mu$m}
\newcommand{\iab}{$i_{\rm AB}$}
\newcommand{\zcl}{$z_{\rm cl}=2.1$}
\newcommand{\sigmacl}{$\sigma_{\rm cl}$}
\newcommand{\ergu}{$\times10^{-17}$~erg~s$^{-1}$~cm$^{-2}$}
\newcommand{\kms}{km~s$^{-1}$}
\newcommand{\Msun}{${\rm M}_{\odot}$}
\newcommand{\Msunyr}{\Msun~yr$^{-1}$}
\newcommand{\Mstar}{${\rm M}_{\star}$}
\newcommand{\Mgas}{${\rm M}_{\rm gas}$}
\newcommand{\fgas}{${\rm f}_{\rm gas}$}
\newcommand{\logMstarMsun}{$\log$(\Mstar/\Msun)}

\newcommand{\Lsun}{L$_{\odot}$}
\newcommand{\lir}{L$_{\rm IR}$}
\newcommand{\lircut}{$2\times10^{11}$~\Lsun}

\newcommand{\Rproj}{R$_{\rm proj}$}
\newcommand{\tdepl}{t$_{\rm depl}$}
\newcommand{\rcirc}{r$_{\rm circ}$}
\newcommand{\rad}{r$_{\rm eff}$}
\newcommand{\drad}{$\Delta[\log$(\rad,\Mstar)$]$}
\newcommand{\sigmaHa}{$\Sigma$(\Halphased)}
\newcommand{\sigmaLIR}{$\Sigma$(\lir)}

\newcommand{\zspec}{$z_{\rm spec}$}
\newcommand{\zphot}{$z_{\rm phot}$}

\newcommand{\Av}{A$_{\rm V}$}
\newcommand{\Rv}{R$_{\rm V}$}
\newcommand{\CalRv}{R$_{\rm V,SB}$}
\newcommand{\Agas}{A(\Halpha)$_{\rm HII}$}
\newcommand{\Ased}{A$_{\rm V,star}$}
\newcommand{\EBV}{{\rm E(B-V)}}
\newcommand{\EBVgas}{\EBV$_{\rm HII}$}
\newcommand{\EBVsed}{\EBV$_{\rm star}$}

\newcommand{\hinverse}{$h^{-1}$}
\newcommand{\rhapsody}{{\sc Rhapsody-G}}
\newcommand{\ramses}{{\sc Ramses}}

\newcommand{\zfire}{{\sc ZFIRE}}
\newcommand{\zfourge}{{\sc ZFOURGE}}

% 90 galaxies at $1.9<z<2.4$ of which 37 are $2.08<z<2.12$
\newcommand{\nfd}{53}
\newcommand{\ncl}{37}
\newcommand{\ntot}{90}
% irbright_fd=14	irbright_cl=7
\newcommand{\nirbright}{21}
\newcommand{\nirfaint}{69}

\title{ZFIRE: Similar Stellar Growth in \Halpha-emitting Cluster and
  Field Galaxies at $z\sim2$}

% Tracking Stellar Growth
%Tracked by \Halpha\ Surface Density and Infrared Luminosity}

\author{Kim-Vy H. Tran\altaffilmark{1,2},
Leo Y. Alcorn\altaffilmark{2},
Glenn G. Kacprzak\altaffilmark{3},
Themiya Nanayakkara\altaffilmark{3},
Caroline Straatman\altaffilmark{4},
Tiantian Yuan\altaffilmark{5},
Michael Cowley\altaffilmark{6,7},
Romeel Dav\'e\altaffilmark{8,9,10},
Karl Glazebrook\altaffilmark{3},
Lisa J. Kewley\altaffilmark{5},
Ivo Labb\'e\altaffilmark{4},
David\'e Martizzi\altaffilmark{11},
%Greg Rudnick\altaffilmark{},
Casey Papovich\altaffilmark{2},
Ryan Quadri\altaffilmark{2},
Lee R. Spitler\altaffilmark{6,7},
Adam Tomczak\altaffilmark{12}
}

\altaffiltext{1}{kimvy.tran@tamu.edu}

\altaffiltext{2}{George P. and Cynthia W. Mitchell Institute for
Fundamental Physics and Astronomy, Department of Physics \& Astronomy,  
Texas A\&M University, College Station, TX 77843}
\altaffiltext{3}{Swinburne University of Technology, Hawthorn, VIC
3122, Australia} 
\altaffiltext{4}{Leiden Observatory, Leiden University, P.O. Box 9513, NL-2300 RA 
Leiden, The Netherlands}
\altaffiltext{5}{Research School of Astronomy and Astrophysics, The
Australian National University, Cotter Road, Weston Creek, ACT 2611,
Australia} 
\altaffiltext{6}{Department of Physics and Astronomy, Faculty of Science and 
Engineering, Macquarie University, Sydney, NSW 2109, Australia} 
\altaffiltext{7}{Australian Astronomical Observatory, PO Box 915, North Ryde, 
NSW 1670, Australia}
\altaffiltext{8}{University of the Western Cape, Bellville, Cape Town,
7535, South Africa}
\altaffiltext{9}{South African Astronomical Observatories, Observatory, Cape
Town, 7925, South Africa}
\altaffiltext{10}{African Institute for Mathematical Sciences,
Muizenberg, Cape Town, 7945, South Africa} 
\altaffiltext{11}{Department of Astronomy, University of California, Berkeley, CA, 
95720}
\altaffiltext{12}{Department of Physics, University of California, Davis, CA, 95616}

\setcounter{footnote}{12}
\begin{abstract}

We compare galaxy scaling relations as a function of environment at
$z\sim2$ with our
\zfire\ survey\footnote{http://zfire.swinburne.edu.au} where we have
measured \Halpha\ fluxes for \ntot\ star-forming galaxies
selected from a mass-limited [\logMstarMsun$>9$] sample based on
\zfourge\footnote{http://zfourge.tamu.edu}.  The cluster galaxies
(\ncl) are part of a confirmed system at $z=2.095$ and the field
galaxies (\nfd) are at $1.9<z<2.4$; all are in the COSMOS legacy
field.  There is no statistical difference between \Halpha-emitting
cluster and field populations when comparing their star formation rate
(SFR), stellar mass (\Mstar), galaxy size (\rad), SFR surface density
[\sigmaHa], and stellar age distributions.  The only difference
  is that at {\it fixed stellar mass}, the \Halpha-emitting cluster
  galaxies are $\log$(\rad)$\sim0.1$ larger than in the field.
  Approximately 19\% of the \Halpha-emitters in the cluster and 26\%
  in the field are IR-luminous (\lir$>$\lircut).  Because the LIRGs in
  our combined sample are $\sim5$ times more massive than the low-IR
  galaxies, their radii are $\sim70$\% larger.  To track stellar
growth, we separate galaxies into those that lie above, on, and below
the \Halpha\ star-forming main sequence (SFMS) using
$\Delta$SFR(\Mstar)$=\pm0.2$~dex.  Galaxies above the SFMS
(starbursts) tend to have higher \Halpha\ SFR surface densities and
younger light-weighted stellar ages compared to galaxies below
the SFMS.  Our results indicate that starbursts (+SFMS) in the cluster
and field at $z\sim2$ are growing their stellar cores.  Lastly, we
compare to the (SFR--\Mstar) relation from \rhapsody\ cluster
simulations and find the predicted slope is nominally consistent with
the observations.  However, the predicted cluster SFRs tend to be too
low by a factor of $\sim2$ which seems to be a common problem for
  simulations across environment.

\end{abstract}


%------------------------------------------------------------
\section{Introduction}

With the discovery and spectroscopic confirmation of galaxy clusters
at $z\sim2$, we have reached the epoch when many massive galaxies in
clusters are still forming a significant fraction of their stars
\citep[$e.g.$][]{tran:10,papovich:10,zeimann:12,brodwin:13,gobat:13,webb:15}.
We can now pinpoint when cluster galaxies begin to diverge from their
field counterparts and thus separate evolution driven by galaxy mass
from that of environment
\citep{peng:10,wetzel:12,quadri:12,muzzin:12,papovich:12,bassett:13}.
At this epoch, measurements of galaxy properties such as stellar mass,
star formation rate, physical size, and metallicity have added
leverage because the cosmic star formation rate density peaks at
$z\sim2$ \citep[see review by][and references therein]{madau:14}.
Observed galaxy scaling relations also test current formation models
\citep[$e.g.$][]{dave:11a,genel:14,tonnesen:14,schaye:15,hahn:15,martizzi:16}.

Particularly useful for measuring galaxy scaling relations at $z\sim2$
are mass-limited surveys because they bridge UV/optical selected
galaxies with the increasing number at $z\gtrsim2$ of dusty
star-forming systems that are IR-luminous but UV faint \citep[see
  reviews by][and references therein]{lutz:14,casey:14}.  Large
imaging surveys have measured sizes and morphologies for galaxies
\citep[$e.g.$][]{wuyts:11,vanderwel:12}, but these studies use
photometric redshifts based on broad-band photometry and are limited
to \logMstarMsun$\gtrsim10$ at $z\sim2$, $i.e.$ just below the
characteristic stellar mass at this epoch \citep{tomczak:14}.  Pushing
to lower stellar masses at $z\sim2$ with more precise star formation
rates requires deep imaging that spans rest-frame UV to near-IR
wavelengths to fully characterize the galaxy Spectral Energy
Distributions (SEDs) and obtain reliable photometric redshifts and
stellar masses \citep{brammer:08,brammer:12,brown:14,forrest:16}.

Here we combine \Halpha\ emission from our \zfire\ survey
\citep{nanayakkara:16} with galaxy properties from the \zfourge\ survey
\citep{straatman:16} with IR luminosities from \spitzer\ to track how
galaxies grow at $z\sim2$.  \zfire\ is a near-IR spectroscopic survey
with MOSFIRE \citep{mclean:12} on Keck I where targets are selected
from \zfourge, an imaging survey that combines deep near-IR
observations taken with the FourStar Imager \citep{persson:13} at the
{\it Magellan Observatory} with public multi-wavelength observations,
$e.g.$ \hubble\ imaging from CANDELS \citep{grogin:11}.
Because \zfire\ is based on \zfourge\ which is mass-complete to
\logMstarMsun$\sim9$ at $z\sim2$ \citep{tomczak:14,straatman:16}, we
can measure galaxy scaling relations for cluster and field galaxies
spanning a wide range in stellar mass.

With spectroscopic redshifts and deep multi-wavelength coverage, we
also are able to compare IR luminous to low-IR galaxies in one of the
deepest mass-limited studies to date.  \citet{swinbank:10} find that
submillimeter galaxies (among the dustiest star-forming systems in the
universe) at $z\sim2$ have similar radii in the rest-frame optical as
``normal'' star-forming field galaxies, but \citet{kartaltepe:12} find
that Ultra Luminous Infra-Red Galaxies (ULIRGs; \lir$>10^{12}$~\Lsun)
at $z\sim2$ have larger radii than typical galaxies.  In contrast,
\citet{rujopakarn:11} find that local ULIRGs have smaller radii than
the star-forming field galaxies.  Because of these conflicting
results, it is still not clear whether the IR-luminous phase for
star-forming galaxies at $z\sim2$ is correlated with size growth.

Alternatively, a more effective approach may be to consider galaxies
in terms of their star formation rate versus stellar mass, $i.e.$ the
Star-Forming Main Sequence \citep[SFMS;][and numerous other
  studies]{noeske:07,whitaker:14,tomczak:16}.  For example,
\citet{wuyts:11} find that galaxies above the SFMS tend to have
smaller effective radii.  By separating galaxies into those above, on,
and below the SFMS, recent studies find that galaxy properties such
as S\'ersic index and gas content correlate with a galaxy's location
relative to the SFMS \citep{genzel:15,whitaker:15}.  However, these
studies use star formation rates based on SED fits to rest-frame UV-IR
observations.  Here we explore these relations using \Halpha\ to
measure the instantaneous SFRs of galaxies at $z\sim2$.

We focus on the COSMOS legacy field where we have identified and
spectroscopically confirmed a galaxy cluster at $z=2.095$
\citep[hereafter the COSMOS cluster;][]{spitler:12,yuan:14}.  We build
on our \zfire\ results comparing the cluster to the field for the gas
phase metallicity-\Mstar\ relation \citep{kacprzak:15,kacprzak:16},
the ionization properties of the Inter-Stellar Medium
\citep[ISM;][]{kewley:16}, and the kinematics and virial masses of
individual galaxies \citep{alcorn:16}.  There are also a number of
luminous infrared sources that are likely dusty star-forming galaxies
in the larger region around the COSMOS cluster \citep{hung:16}.

We use a Chabrier Initial Mass Function and AB magnitudes throughout
our analysis.  We assume $\Omega_{\rm m}=0.3$, $\Omega_{\Lambda}$=0.7,
and $H_0=70$~\kms~Mpc$^{-1}$.  At $z=2$, the angular scale is
$1''=8.37$~kpc.

%------------------------------------------------------------

\section{Observations}

\subsection{\zfourge\ Catalog}

To select spectroscopic targets in the COSMOS field, we use the
\zfourge\ catalog that provides high accuracy photometric redshifts
based on multi-filter ground and space-based imaging
\citep{straatman:16}.  \zfourge\ uses EAZY
\citep{brammer:08,brammer:12} to first determine photometric redshifts
by fitting Spectral Energy Distributions, and then FAST
\citep{kriek:09a} to measure rest-frame colors, stellar masses,
stellar attenuation, and specific star formation rates for a given SF
history.  We use a \citet{chabrier:03} initial stellar mass function,
constant solar metallicity, and exponentially declining star formation
rate ($\tau=$10 Myr to 10 Gyr).  For a detailed description of the
\zfourge\ survey and catalogs, we refer the reader to
\citet{straatman:16}.

An advantage of using the deep \zfourge\ catalog is that we can
optimize the target selection to MOSFIRE, specifically by selecting
star-forming galaxies as identified by their UVJ colors
\citep[$e.g.$][]{wuyts:07,williams:09}.  Because the \zfourge\ catalog
reaches Fourstar/Ks$=25.3$ mag and fits the SEDs from the UV to MIR
\citep{straatman:16}, we are able to obtain MOSFIRE spectroscopy for
objects with stellar masses down to \logMstarMsun$\sim9$ at $z\sim2$
\citep{nanayakkara:16}.  Our analysis focuses on the star-forming
galaxies, thus we remove AGN identified in \citet{cowley:16}'s
multi-wavelength catalog.

\subsection{Keck/MOSFIRE Spectroscopy}

We refer the reader to \citet{nanayakkara:16} and \citet{tran:15} for
an extensive description of our Keck/MOSFIRE data reduction and
analysis.  To briefly summarize, the spectroscopy was obtained on
observing runs in December 2013 and February 2014.  A total of eight
slit-masks were observed in the K-band with total integration time of
2 hours each.  The K-band wavelength range is $1.93-2.38\mu$m and the
spectral dispersion of 2.17~\AA/pixel. We also observed two masks in
H-band covering $1.46-1.81\mu$m with a spectral dispersion
1.63~\AA/pixel.

To reduce the MOSFIRE spectroscopy, we use the publicly available data
reduction pipeline (DRP) developed by the instrument
team\footnote{https://github.com/Mosfire-DataReductionPipeline/MosfireDRP}.
We then apply custom IDL routines to correct the reduced 2D spectra
for telluric absorption, spectro-photometrically calibrate by
anchoring to the well-calibrated photometry, and extract the 1D
spectra with assocated $1\sigma$ error spectra
\citep[see][]{nanayakkara:16}.  We reach a line-flux of
$\sim0.3$\ergu\ \citep[$5\sigma$;][]{nanayakkara:16}.  In our
analysis, we select galaxies with \Halpha\ redshifts of $1.9<z<2.4$,
$i.e.$ corresponding to the K-band wavelength range, and exclude AGN
(3 in cluster, 6 in field) identified by \citet{cowley:16}.

As reported in \citet{nanayakkara:16}, our success rate of detecting
\Halpha\ emission at $S/N>5$ in the K-band is $\sim73$\% and the
redshift distribution of the \Halpha-detected galaxies is the same as
the expected redshift probability distribution from \zfourge\ (see
their Fig.~6).  A higher success rate is nearly impossible given the
number of strong sky-lines within the K-band.  We also confirm that
the \zfire\ galaxies are not biased in stellar mass compared to the
\zfourge\ photometric sample \citep[see their \S3.3 \&
Fig.~8]{nanayakkara:16}. 

\begin{figure}
\figurenum{1}
\plotone{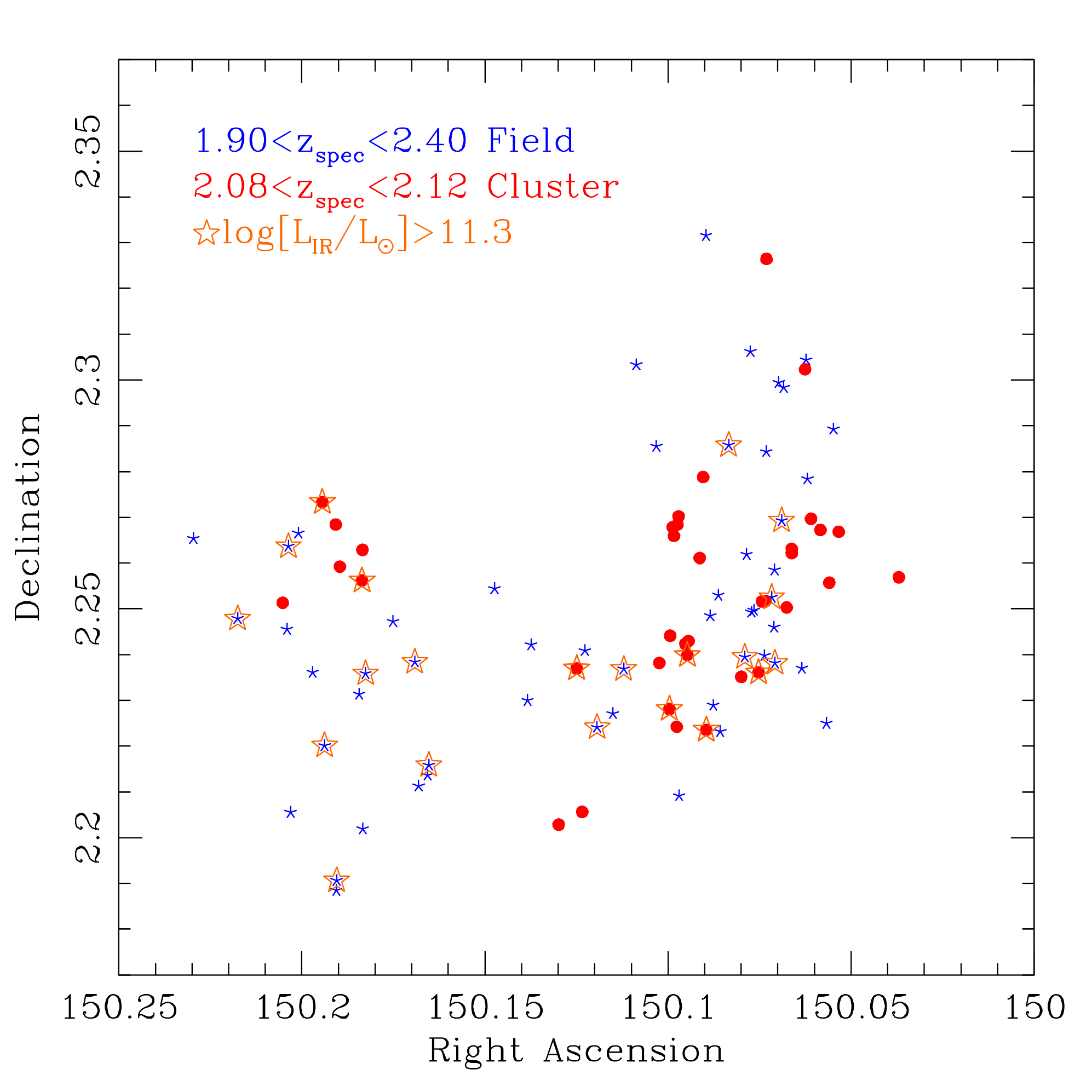}
\caption{Spatial distribution of \Halpha-emitting cluster (filled
  circles; \ncl) and field (crosses; \nfd) galaxies at $z\sim2$ in the
  COSMOS legacy field.  Galaxies with total IR luminosities
  \lir$>$\lircut\ as measured using \spitzer/\mipsmu\ ($3\sigma$
  detection) are shown as open stars (\nirbright).  Active Galactic
  Nuclei (AGN) are excluded using the AGN catalog by
  \citet{cowley:16}.  The fraction of IR-luminous galaxies is the same
  in the field and the cluster ($\sim20-25$\%).
\label{fig:xyplot}}
\end{figure}

Figure~\ref{fig:xyplot} shows the spatial distribution of our
\ncl\ cluster and \nfd\ field galaxies at $z\sim2$.  Cluster members
have spectroscopic redshifts of $2.08<$\zspec$<2.12$
\citep{yuan:14,nanayakkara:16} and field galaxies have \zspec\ of
$(1.97-2.06)$ and $(2.13-2.31)$.  We consider only galaxies with
\zspec\ quality flag Q$_z=3$.  To test if our field sample is
contaminated by cluster galaxies, we also apply a more stringent
redshift selection of $(1.97-2.03)$ and $(2.17-2.31)$ which
corresponds to $>8$ times the cluster's velocity dispersion from the
cluster redshift \citep[$\sigma_{1D}=552$~\kms;][]{yuan:14}.  We
confirm that using the more conservative redshift range for the field
does not change our following results.

We note that our study focuses on cluster and field galaxies at
$z\sim2$ identified by their \Halpha\ emission, thus we cannot
confidently measure the relative fraction of star-forming to all
galaxies across environment with the current dataset.

\subsection{Measuring Galaxy Sizes \& Morphologies}\label{sec:galfit}

We use {\sc GALFIT} \citep{peng:10} to measure S\'ersic indices,
effective radii, axis ratios, and position angles for the
spectroscopically confirmed galaxies in COSMOS using \hubble\ imaging
taken with WFC3/F160W.  Most of these galaxies are in the
\citet{vanderwel:12} morphological catalog which spans a wide redshift
range.  However, we choose to measure independently the galaxy sizes
and morphologies to optimize the fits for our galaxies at $z\sim2$.

\begin{figure}
\figurenum{2}
\plotone{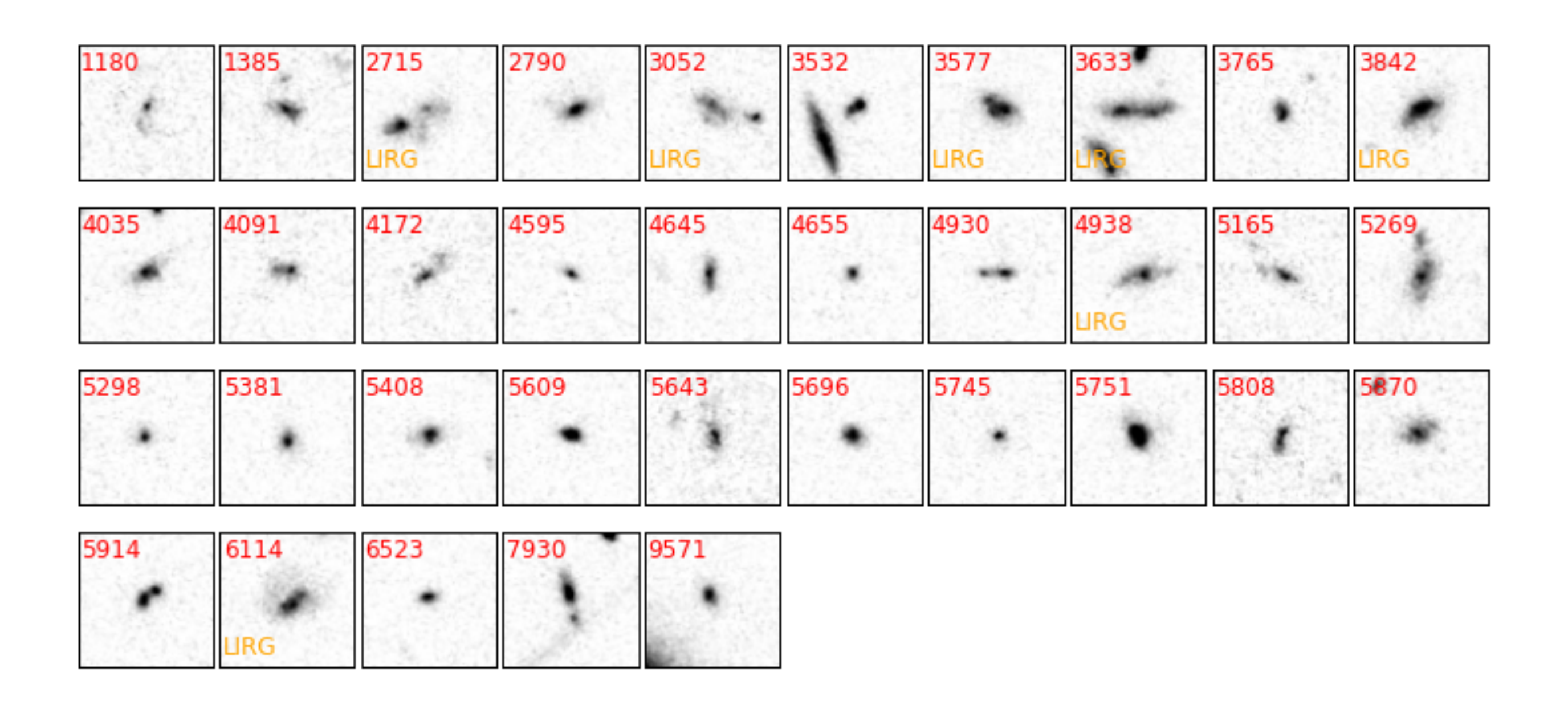}
\caption{HST images ($4''\times4''$) generated by summing F125W,
  F140W, and F160W for \Halpha-emitting cluster galaxies
  ($2.08<$\zspec$<2.12$); S\'ersic indices and effective radii are
  measured using GALFIT for 35 of \ncl\ members.  Galaxies are labeled
  with their \zfire\ IDs and IR-luminous galaxies (\lir$>$\lircut) are
  noted as LIRGs.
\label{fig:cluster_hst}}
\end{figure}

\begin{figure}
\figurenum{3}
\plotone{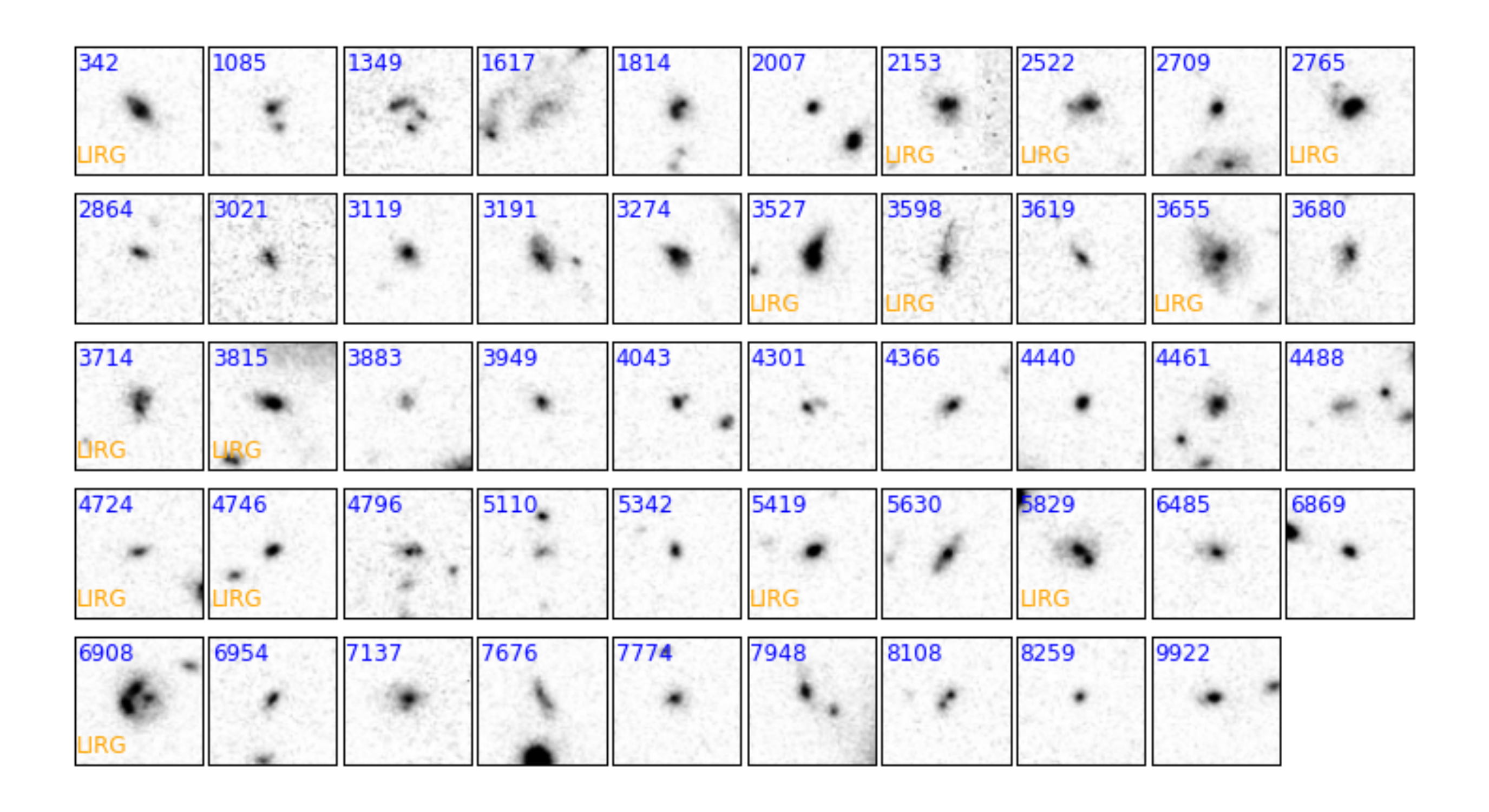}
\caption{HST images ($4''\times4''$) generated by summing F125W,
  F140W, and F160W for \Halpha-emitting field galaxies at $z\sim2$
  ($1.9<$\zspec$<2.4$); S\'ersic indices and effective radii are
  measured using GALFIT for 49 of \nfd\ field galaxies.  Galaxies are
  labeled with their \zfire\ IDs and IR-luminous galaxies
  (\lir$>$\lircut) are noted as LIRGs.
\label{fig:field_hst}}
\end{figure}

Of the \ntot\ galaxies in our \Halpha-emitting sample, we measure
effective radii along the major axis and S\'ersic indices for 83
(35 cluster, 48 field); see Figs.~\ref{fig:cluster_hst} \&
\ref{fig:field_hst} for galaxy images and Table~1 for galaxy
properties.  Seven of the galaxies could not be fit because of
  contamination due to diffraction spikes from nearby stars or
  incomplete F160W imaging \citep[see][]{skelton:14}.  We include a
quality flag on the GALFIT results and identify 12 galaxies with fits
that have large residuals due to, $e.g.$ being mergers
\citep[see][]{alcorn:16}.  We confirm that excluding these 12 galaxies
does not change our general results and so we use the effective radii
measured for all 83 galaxies in our analysis.

Following \citet{vanderwel:14}, we use the effective radius to
characterize size because \rad\ is more appropriate than a
circularlized radius for galaxies spanning the range in axis ratios.
We confirm that using \rcirc\ instead of \rad\ does not change the
following results except for shifting the size distribution of the
entire galaxy sample to smaller sizes.  The trends in the scaling
relations that depend on galaxy size, $e.g.$ comparing cluster to
field and galaxies relative to the SFMS, are robust.

\subsection{Dust-corrected \Halpha\ Star Formation Rates}\label{sec:dust}

To use \Halpha\ line emission as a measure of star formation rate
(SFR), we need to correct for dust attenuation.  Although determining
the internal extinction using the Balmer decrement is preferred, we
have \Hbeta\ for only a small subset.  Thus we must rely on the
stellar attenuation \Ased\ measured by FAST which assumes \Rv=4.05
\citep[starburst attenuation curve;][]{calzetti:00}\footnote{The
  starburst (SB) attenuation curve is commonly referred to as the
  Calzetti law and is appropriate for continuum measurements.  We use
  ``starburst'' as requested by D. Calzetti.}.  For more extensive
results on stellar vs. Balmer-derived attenuation and star formation
rates, we refer the reader to \citet{price:14} and \citet{reddy:15}.

Following \citet{tran:15} \citep[see also][]{steidel:14}, the
\Halpha\ line fluxes are corrected using the nebular attenuation curve
from \citet{cardelli:89} with \Rv=3.1

\begin{equation}
{\rm A}({\rm H}\alpha)_{\rm HII} = 2.53 \times {\rm E(B-V)}_{\rm HII}.
\end{equation}

We use the observed stellar to nebular attenuation ratio of
\EBVsed$=0.44\times$\EBVgas\ \citep{calzetti:00} and the color excess
${\rm E(B-V)}_{\rm star}$ which is the stellar attenuation
  \Ased\ measured by FAST divided by \Rv=4.05.  Combining these
  factors, we have

\begin{equation}
{\rm A}({\rm H}\alpha)_{\rm HII} = 5.75 \times {\rm E(B-V)}_{\rm
star}\label{eq:AHa}
\end{equation}

which we use to correct all of the \Halpha\ fluxes for attenuation.
Recent work by \citet{reddy:15} suggests that the \EBVsed\ to
\EBVgas\ ratio may depend on stellar mass at $z\sim2$, but there is
significant scatter in the fitted relation.  We stress that such a
correction would not change our results because we use the same method
to measure \Halpha-SFRs for all the galaxies in our study and compare
internally.

We determine the corresponding star formation rates using the relation
from \citet{hao:11}:

\begin{equation}
\log[{\rm SFR}({\rm H}\alpha_{\rm star})] = \log[{\rm L}({\rm H}\alpha_{\rm star})] - 
41.27
\end{equation}

This relation assumes a Kroupa IMF
\citep[$0.1-100$~\Msun;][]{kroupa:01}, but the relation for a Chabrier
IMF is virtually identical (difference of 0.05).  Note that
$\log$[SFR(\Halphased)] values determined with the \citet{hao:11}
relation are $0.17$~dex lower than when using the \citet{kennicutt:98}
relation.

\subsection{\Halphased-SFR Surface Densities}

With the \Halphased\ SFRs and galaxy sizes as measured by their
effective radii (\rad), we can then determine the SFR surface density:

\begin{equation}
\Sigma({\rm H}\alpha_{\rm star}) =
\frac{ {\rm SFR(H}\alpha_{\rm star}) }{2 \pi \times {\rm r}_{\rm eff}^2}
\end{equation}

\begin{figure}
\figurenum{4}
\plotone{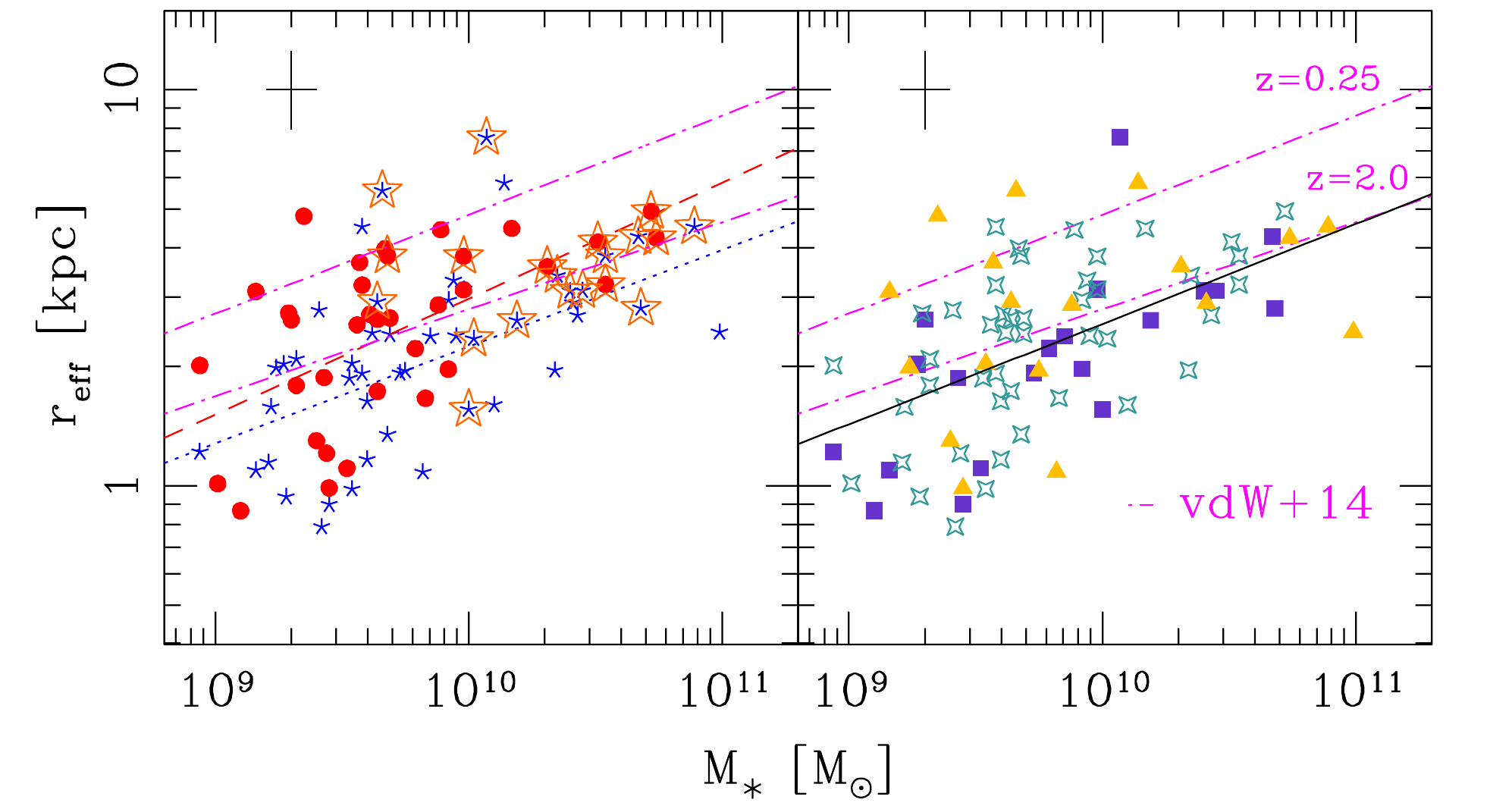}
\caption{We measure the effective radii (\rad) using \hubble\ imaging
  taken with WFC3/F160W. {\it Left:} The galaxy size-stellar mass
  relation for our combined sample is consistent with the fit to
  star-forming galaxies at $z\sim2$ measured using photometric
  redshifts by CANDELS and clearly offset from the relation at
  $z=0.25$ \cite[pink dash-dot curves;][]{vanderwel:14}.  We find no
  significant difference between the size-mass relation for
  \Halpha-emitting cluster (red dashed) and field (blue dotted)
  galaxies at $z\sim2$.  {\it Right:} The \rad-\Mstar\ relations for
  galaxies on (open crosses) and below (filled triangles) the
  \Halpha\ star-forming main sequence (SFMS; see
  Fig.~\ref{fig:sfr_smass}) are consistent with CANDELS, but the
  galaxies with elevated SFRs (filled squares) have smaller radii at a
  given stellar mass. For reference, the black line is the
    ($2\sigma$ clipped) least squares fit to our combined sample.
\label{fig:rad_smass}}
\end{figure}

Note that most of the cluster and field galaxies have effective radii
of \rad$\sim0.35''$ (Fig.~\ref{fig:rad_smass}) which is comparable to
the slit-width of $0.7''$.  

It is possible that by using \rad\ measured with WFC/F160W imaging, we
are overestimating \sigmaHa.  \citet{forster:11} find that the
\Halpha\ sizes of six $z\sim2$ galaxies are comparable to their
rest-frame continuum sizes as measured with IFU and HST observations.
However, \citet{nelson:16a} show that at $z\sim1$, continuum-based
sizes tend to be smaller than \Halpha-based sizes for star-forming
galaxies with \logMstarMsun$\gtrsim10$.  While correcting for a
possible dependence of \Halpha-size on galaxy mass would shift
\sigmaHa\ to lower values, it would not change our overall conclusions
based on comparing the different galaxy populations.

Note that with our current single-slit observations, we cannot address
a possible environmental dependence of \Halpha-disks.  Galaxies in the
Virgo cluster are known to have truncated \Halpha-disks compared to
the field \citep{kenney:99,koopmann:04}, thus not accounting for disk
truncation in the cluster galaxies may lead to over-estimating their
total \Halphased-SFRs and consequently \sigmaHa.  Future deep IFU
observations with the next generation of large telescopes should be
able to test for \Halpha-disk truncation in these $z\sim2$ galaxies.

\subsection{IR Luminosities from Spitzer/MIPS}

Summarizing from \citet{tomczak:16}, IR luminosities are determined
from Spitzer/MIPS observations at \mipsmu\ (GOODS-S: PI M. Dickinson,
COSMOS: PI N. Scoville, UDS: PI J.  Dunlop) which have $1\sigma$
uncertainties of 10.3 $\mu$Jy in COSMOS.  We measure the
\mipsmu\ fluxes within $3.5''$ apertures and use the custom code
MOPHONGO \citep[written by I.~Labb\'e; see][]{labbe:06,wuyts:07} to
deblend fluxes from multiple sources. The \citet{wuyts:08} templates
are fit to the SEDs using the \Halpha-redshifts to determine
integrated $8-1000$~$\mu$m fluxes; we refer the reader to
\citet{tomczak:16} for a full description of the IR measurements.

For galaxies at $z\sim2$, the $3\sigma$ \lir\ detection limit is
\lircut, $i.e.$ all our \lir\ galaxies are Luminious Infra-Red
Galaxies\footnote{Note that our \lir\ detection limit is higher than
  the LIRG threshold of $10^{11}$~\Lsun\ \citep[see review
    by][]{sanders:96}, $i.e.$ we do not detect LIRGs with
  ($10^{11}$~\Lsun$<$\lir$<$\lircut).  Thus some of our low-IR
  galaxies may still technically be LIRGs}.  Figure~\ref{fig:xyplot}
shows the spatial distribution of IR-luminous cluster and field
galaxies.  In our analysis, we use IR-based luminosities and
\Halphased\ SFRs.  We note that \lir\ detection thresholds at
  $z>1$ correspond to star formation rates that are much higher than
  UV-based SFRs.  Thus comparing, e.g. \Halphased-SFR to a combined
  (IR+UV) SFR instead of an \lir-only SFR does not change our
  results. 

%------------------------------------------------------------

\section{Results}

\subsection{A Population of IR-Luminous Galaxies}\label{sec:IRfrac}

A remarkable 19\% (7/\ncl) of \Halpha-emitting cluster galaxies
  at $z\sim2$ have \lir$>$\lircut.  Within errors, this fraction of
  IR-luminous cluster galaxies is comparable to the field (26\%,
  14/\nfd; Fig.~\ref{fig:xyplot}).  \citet{saintonge:08} showed using
\mipsmu\ observations of $\sim1500$ spectroscopically confirmed
cluster galaxies that the fraction of IR members increases with
redshift, but this was limited to galaxy clusters at $0<z<1$.  More
recent studies using the {\it Herschel Space Observatory} have
detected IR sources in galaxy clusters at $z>1$
\citep{popesso:12,santos:14}, but far-IR observations can only detect
a handful of the most IR-luminous systems with star formation rates
$>100$~\Msunyr.  Our survey is the first to spectroscopically confirm
the high fraction of LIRGs in galaxy clusters at $z\sim2$ \citep[see
  also][]{hung:16}.

\subsection{Comparing Star Formation Rates}\label{sec:sfr_compare}

\begin{figure}
\figurenum{5}
\plotone{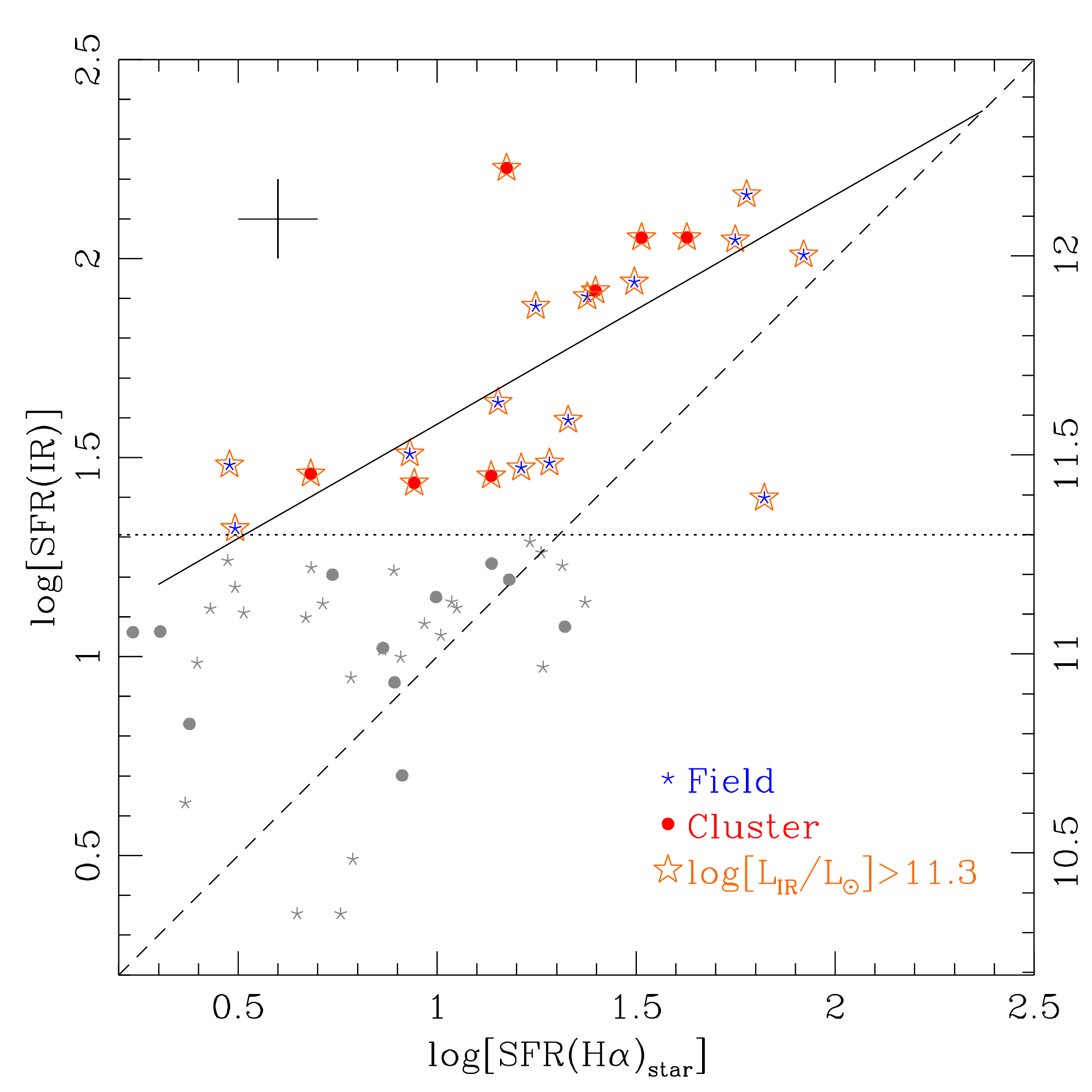}
\caption{A Spearman rank test confirms that for the 21 galaxies with
  SFR(\Halphased)$>$\Halphacut\ and \lir$>$\lircut\ (horizontal dotted
  line), their SFRs based on these two tracers are correlated
  ($>2\sigma$ confidence).  The solid line shows the best fit least
  squares ($2\sigma$-clipped) and the dashed diagonal line is parity;
  the cross in the upper left shows a representative log error of
  $\pm0.1$~dex. Galaxies with \lir$<$\lircut\ are shown in gray
    and have \lir\ errors larger than the representative value.
  There is no evidence of environmental dependence: K-S tests confirm
  that the \Halphased\ and \lir\ star formation rates have the same
  parent populations for cluster and field galaxies.  The same is true
  if we compare the combined (IR+UV) star formation rate to
  \Halphased\ values.  However, SFRs based on \Halphased\ are
  systematically lower than those from \lir.
\label{fig:Ha_lir}}
\end{figure}

\subsubsection{Cluster vs. Field}

We find no evidence of different correlations between \Halpha\ and
\lir\ when considering the cluster and field samples separately
(Fig.~\ref{fig:Ha_lir}; Table~1).  For the 14 field and 7 cluster
galaxies with \lir$>$\lircut, a K-S test measures a p-value of 0.13,
$i.e.$ the statistical likelihood of the cluster and field populations
being drawn from different parent populations is low.  The average
$\log$(\lir) per galaxy is comparable: $11.7\pm0.3$ in the field
versus $11.8\pm0.3$ in the cluster.  This is true also when selecting
instead by SFR(\Halphased)$>$\Halphacut: the field (52) and cluster
(34) populations have the same median $\log$[SFR(\Halphased)] of
$0.9\pm0.3$.  Note that K-S tests confirm the \Halpha-emitting
galaxies in the cluster and field are drawn from the same parent
population in terms of their stellar mass and Specific Star Formation
Rate (SSFR$=$SFR/\Mstar).

\subsubsection{\Halpha\ vs. \lir}

For galaxies with both \Halphased$>$\Halphacut\ and
\lir$>$\lircut\ (21), a Spearman rank test confirms a positive
correlation ($>2\sigma$) between SFRs based on these two tracers
\citep[Fig.~\ref{fig:Ha_lir}, Table~1; see
  also][]{ibar:13,shivaei:16}.  However, the dust-corrected
\Halphased\ SFRs are systematically lower than \lir\ SFRs by
$\sim0.5$~dex, $i.e.$ by nearly a factor of 3.  This is driven mostly
by a combination of using the \citet{hao:11} relation for converting
\Halpha\ luminosities to SFRs instead of, $e.g.$ \citet{kennicutt:98},
and by choice of dust law.  We confirm that comparing \Halphased\ to a
combined (IR+UV) star formation rate does not change our results.

We measure a scatter of $\sigma\sim0.33$~dex in \Halphased-\lir\ SFRs
which is larger than $\sigma\sim0.22$~dex measured recently by
\citet{shivaei:16} for 17 galaxies at $z\sim2$.  However, their
analysis focuses on galaxies with SFRs $>10$~\Msunyr\ while we push to
\Halphased\ SFRs of $\sim$\Halphacut.  From Fig.~\ref{fig:Ha_lir}, the
discrepancy between \Halphased\ and \lir\ SFRs decreases at higher
values.

\subsection{\Halpha\ Star-Forming Main Sequence at
  $z\sim2$}\label{sec:sfr_smass} 

\begin{figure}
\figurenum{6}
\plotone{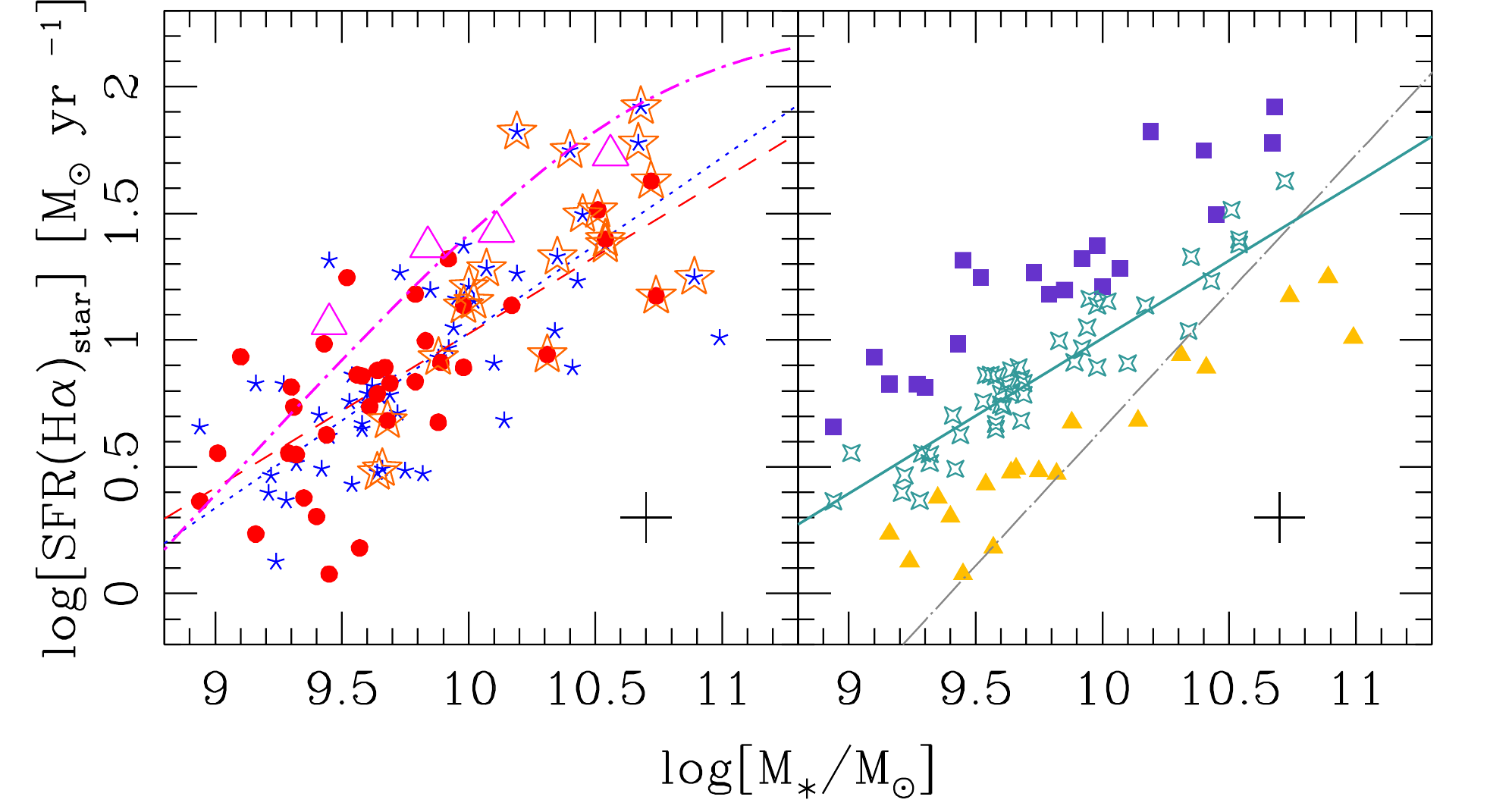}
\caption{{\it Left:} At $z\sim2$, the COSMOS cluster (red filled
  circles) and field (blue line stars) galaxies follow identical
  relations between stellar mass and \Halphased\ star formation rate;
  $2\sigma$-clipped least-squares fits are shown by red dashed and
  blue dotted lines, respectively.  The cross in the lower right shows
  a representative log error of $\pm0.1$~dex.  Both fits are
  consistent with the shape of the SFR-\Mstar\ relation measured by
  \zfourge\ for star-forming field galaxies at $z\sim2$ using
  photometric redshifts \citep[pink curve][]{tomczak:16} as well as
  the mass-binned sample from MOSDEF for \Halpha-selected field
  galaxies at $z\sim2$ \citep[open triangles;][]{sanders:15}; because
  we use \citet{hao:11} to convert \Halpha\ luminosity to SFR, we are
  offset in $\log$[SFR(\Halphased)] from both \zfourge\ and MOSDEF.
  The more massive galaxies [\logMstarMsun$>10$] tend to be
  IR-luminous (\lir$>$\lircut; open orange stars), $i.e.$ they are
  LIRGs.  {\it Right:} We fit the \Halpha\ star-forming main sequence
  (hereafter SFMS) using our combined cluster and field sample (cyan
  line).  In our analysis, we consider star-forming galaxies that lie
  above (+SFMS; purple filled squares), on (=SFMS; cyan open crosses),
  and below (--SFMS; yellow filled triangles) the \Halpha\ SFMS .
  Also shown is the predicted SFMS relation at $z\sim2$ from
  \rhapsody, a high-resolution AMR simulation of galaxy clusters
  \citep[gray long dash-dot line;][]{martizzi:16}.
\label{fig:sfr_smass}}
\end{figure}

Using deep multi-wavelength imaging, the relation between star
formation rate and stellar mass is now measured to $z\sim3$ for
thousands of galaxies down to \logMstarMsun$\sim9$ \citep[e.g.][see
  Fig.~\ref{fig:sfr_smass}]{whitaker:12,tomczak:16}.  However, the
SFRs and stellar masses derived by fitting Spectral Energy
Distributions (SEDs) to multi-wavelength imaging can be degenerate.
Measurements of \Halpha\ fluxes are a more accurate tracer of the
instantaneous SFR than fitting SEDs to photometry
  \citep{kennicutt:12}, but are restricted to a smaller sample of
galaxies due to the observational challenge of measuring \Halpha\ at
$z\sim2$.

Combining SFRs based on \Halphased\ fluxes and stellar masses derived
from SED fitting, we fit the SFR-\Mstar\ relation using a
($2\sigma$-clipped) least-squares for the field and cluster
populations separately.  Note that the field and cluster galaxies span
the full range in both stellar mass and
\Halphased-SFR\ (Fig.~\ref{fig:sfr_smass}).  The cluster and field
galaxies at $z\sim2$ have the same increasing SFR-\Mstar\ relation:

\begin{equation}
\log[{\rm SFR(H}\alpha_{\rm star},{\rm Field)}] = 0.69\log ( {\rm M}_{\star} ) - 5.82
\end{equation}

\begin{equation}
\log[{\rm SFR(H}\alpha_{\rm star},{\rm Cluster)}] = 0.62\log ( {\rm M}_{\star} ) - 
5.15
\end{equation}

\begin{equation}
\log[{\rm SFR(H}\alpha_{\rm star},{\rm All)}] = 0.61\log ( {\rm M}_{\star} ) - 
5.11\label{eq:sfms}
\end{equation}

where SFR is in \Msunyr\ and \Mstar\ is in \Msun.  The rms error on
the fitted slopes is $\sim0.2$, and separate 1D K-S tests confirm that
the stellar mass and SFR distributions of our cluster and field
populations are similar.  A possible concern is that our field sample
could be contaminated by cluster members, but we confirm that applying
a more stringent redshift cut of $>8\sigma_{\rm 1D}$ to select field
galaxies does not change our results.

Our measurements are consistent with recent results, $e.g.$ from
\zfourge\ \citep[SED fitting of UV-MIR;][]{tomczak:16} and MOSDEF
\citep[\Halpha;][]{sanders:15}, and span similar ranges in stellar
mass and star formation rate.  However, our \Halphased\ SFRs are
lower.  This offset is mostly likely due to differences in the
relation used to convert \Halpha\ luminosities to SFRs, $e.g.$
\citet{hao:11} vs. \citet{kennicutt:98}, and choice of dust law.
Accounting for both these effects increases $\log$[SFR(\Halphased)] by
$\sim0.3$~dex which brings our SFMS into agreement with \zfourge\ and
MOSDEF.  These systematic differences in SFRs due to using
  different conversion relations and dust laws highlights the need to
  identify a more robust method of measuring SFRs at $z>1$
  \citep[e.g.][]{reddy:15,shivaei:16}.

In our analysis, we also compare star-forming galaxies that lie above,
on, and below the star-forming main sequence (hereafter SFMS) as
measured by \Halpha\ emission.  Using the best-fit to the combined
cluster and field sample (Eq.~\ref{eq:sfms}), we calculate a galaxy's
offset from the \Halpha\ SFMS given its stellar mass.  Because the
typical scatter in the \Halpha\ SFMS is $\sim0.2$ dex, we use
$\Delta$SFR(\Mstar)$=0.2$~dex to separate galaxies into those above
(20), on (45), and below (18) the SFMS.  Galaxies in these three
classes (+SFMS, =SFMS, --SFMS) span the full range in stellar mass
(Fig.~\ref{fig:sfr_smass}, right).

The LIRGs also span the full range in stellar mass and \Halphased-SFR
for both field and cluster galaxies, and the most massive galaxies
[\logMstarMsun$\gtrsim10$] tend to be LIRGs (Fig.~\ref{fig:sfr_smass},
left).  The LIRGs at $z\sim2$ follow the same trend of increasing
\Halphased-SFR with stellar mass (Fig.~\ref{fig:sfr_smass}; slope
$\sim0.80$), a somewhat surprising result given the large scatter when
comparing SFRs derived from \Halphased\ to \lir\ (see
\S\ref{sec:sfr_compare}).  LIRGs lie above, on, and below the
star-forming main sequence as defined by their \Halphased-SFRs\
(Fig.~\ref{fig:sfr_smass}, right).

\subsection{Galaxy Size-Stellar Mass Relation}\label{sec:sizes}

How galaxy size correlates with stellar mass depends on galaxy type,
$e.g.$ quiescent galaxies with S\'ersic indices of $n\sim4$ tend to be
smaller at a given stellar mass compared star-forming galaxies with
$n\sim1$ \citep{shen:03}.  With a limited spectroscopic sample of
galaxies, \citet{law:12} showed that the galaxy size-mass relation
evolves with redshift.  Most recently, \citet{vanderwel:14} used high
resolution imaging from \hubble\ and photometric redshifts for
$\sim31,000$ galaxies to measure how the \rad--\Mstar\ relation of
star-forming galaxies has evolved since $z\sim3$.

We measure S\'ersic indices and effective radii for 83 of the
\ntot\ galaxies in our sample (see \S\ref{sec:galfit} and Table~1).
We find that our \Halpha-emitting $z\sim2$ galaxies follow the same
trend of increasing galaxy size with stellar mass measured by
\citet{vanderwel:14} for galaxies at this epoch
(Fig.~\ref{fig:rad_smass}).  Most of our fitted galaxies (71 of 83)
have S\'ersic indices of $n\leq2$, and most (80 of 83) have effective
radii of $0.7<$\rad$<5$~kpc (Fig.~\ref{fig:hist_rad}).  

\subsubsection{Cluster vs. Field}

\begin{figure}
\figurenum{7}
\plotone{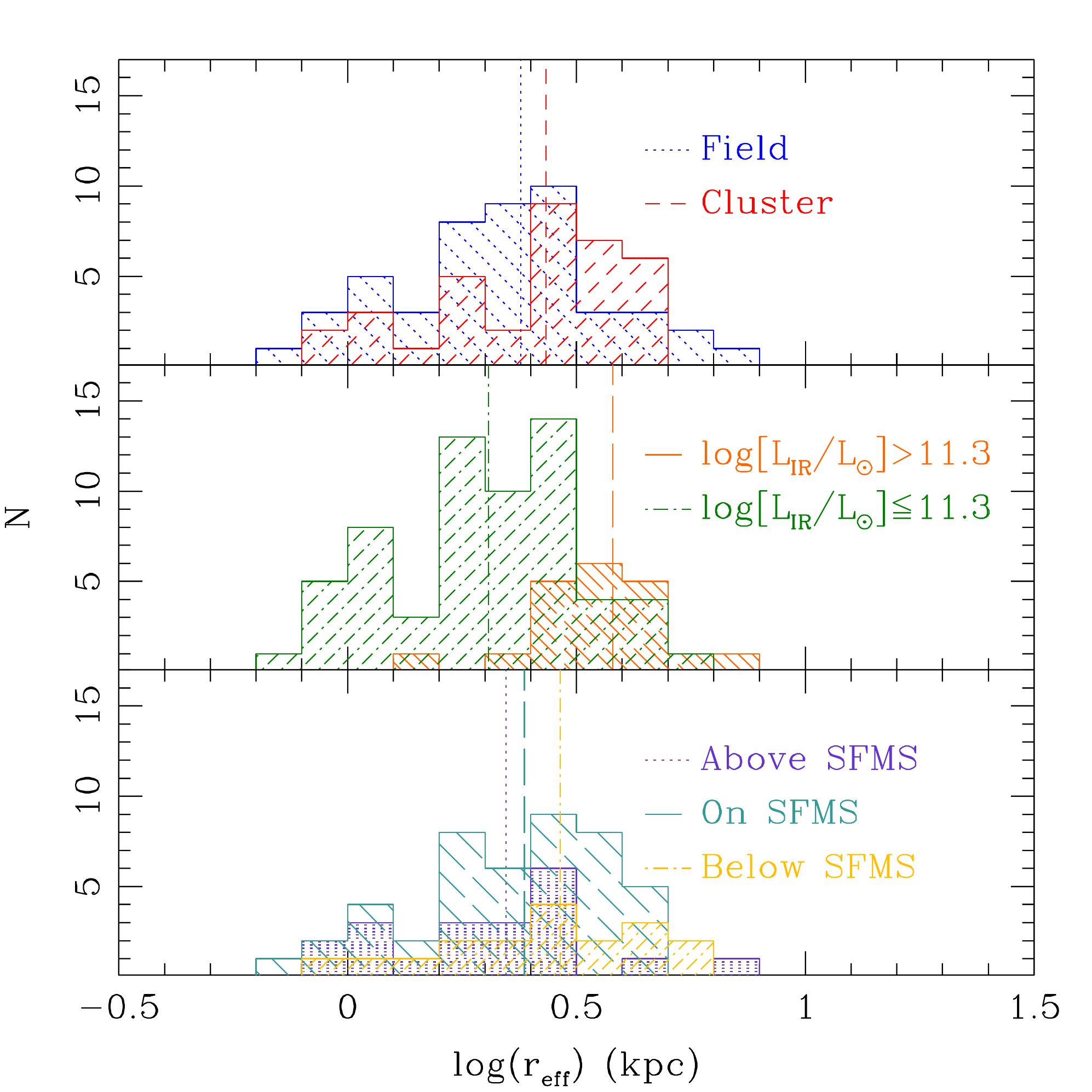}
\caption{{\it Top:} \Halpha-selected cluster and field galaxies at
  $z\sim2$ have the same size distribution as measured by the
  effective radius (\rad); medians are shown as vertical lines.  {\it
    Middle:} However, in the combined sample, the IR-luminous galaxies
  (\lir$>$\lircut) tend to be $\sim0.25$ dex larger ($\sim70$\% larger
  in linear space) than the low-IR galaxies.  A K-S test confirms at
  $>3\sigma$ significance that the LIRGs and low-IR galaxies have
  different size distributions.  The LIRGs also tend to be more
  massive (see Fig.~\ref{fig:sfr_smass}).  {\it Bottom:} Galaxies
  above, on, and below the \Halpha\ star-forming main sequence span a
  similar range in galaxy size, but +SFMS galaxies tend to have
  smaller \rad\ at a given stellar mass compared to --SFMS
  galaxies (Fig.~\ref{fig:rad_smass}).
\label{fig:hist_rad}}
\end{figure}

We find no difference in the galaxy size-stellar mass relation with
environment for \Halpha-emitting galaxies.  The cluster and field
populations have the same size distributions with similar average
effective radii of \rad$\sim2.5\pm0.2$~kpc and
\rad$\sim2.2\pm0.2$~kpc, respectively (Fig.~\ref{fig:hist_rad}).
Least-squares fits to the \rad--\Mstar\ distribution for the cluster
and field populations confirms that, within the errors, the
least-squares fits agree with the \citet{vanderwel:14} size-mass
relation.

The astute reader may notice possible conflict with our results in
\citet{allen:15} reporting that star-forming cluster galaxies are
$\sim12$\% larger than in the field.  However, we do find evidence
that at {\it fixed stellar mass}, our cluster galaxies are
$\sim0.1$~dex larger which is consistent with \citet{allen:15}.  We
refer to \S\ref{sec:drad} below for details. 

\subsubsection{IR-Luminous Galaxies}

IR-luminous galaxies (LIRGs) have different physical size and stellar
mass distributions relative to the low-IR population.  A K-S test of
the size distributions (Fig.~\ref{fig:hist_rad}) confirms with
$>3\sigma$ significance that the LIRGs are larger with a median
\rad$\sim3.8$~kpc compared to $\sim2.0$~kpc for the low-IR galaxies
(typical errors for both are $\sim0.3$~kpc).  LIRGs also are $\sim5$
times more massive with \logMstarMsun$\sim10.4$ compared to $\sim9.6$
for the low-IR galaxies (Figs.~\ref{fig:rad_smass} \&
\ref{fig:sfr_smass}).  Even if we consider only galaxies with
  \logMstarMsun$>9.6$, LIRGs and low-IR galaxies have statistically
  different absolute \rad\ distributions.

The size difference between our LIRGs and the low-IR galaxies at
$z\sim2$ seem to be in conflict with \citet{swinbank:10} who, using
\hubble/WFC3/F160W imaging of 25 submm galaxies at $\bar{z}\sim2.1$,
find their submm galaxies have the same sizes as field galaxies at
$1<z<3.5$ (both have typical half-light radii of $\sim2.5-2.8$~kpc).
We find our LIRGs are typically $\sim70$\% larger than the low-IR
population \citep[see also][]{kartaltepe:12}.  This discrepancy is
likely due to our IR comparison being based on a mass-selected sample
that identifies LIRGs to \logMstarMsun$\sim9.5$
(Fig.~\ref{fig:sfr_smass}) while \citet{swinbank:10} is limited to
galaxies with \logMstarMsun$>10$, $i.e.$ galaxies that are large
regardless of their \lir\ emission because they are massive.

\subsubsection{Above, On, \& Below the \Halpha\ Star-Forming Main Sequence}\label{sec:sfms}

Galaxies above, on, and below the \Halpha\ star-forming main sequence
(SFMS; see Fig.~\ref{fig:sfr_smass}, right) also follow the same
general trend of increasing galaxy size with stellar mass
(Fig.~\ref{fig:rad_smass}, right).  K-S tests confirm that the size
distributions for all three groups are likely drawn from the same
parent population.

One concern in using \Halpha\ SFRs obtained with slit spectroscopy
is that we are biased towards compact star-forming galaxies, $e.g.$
significant slit losses in the spectroscopic flux measurements will
cause smaller galaxies to appear to have higher \Halpha-SFRs
compared to larger galaxies.  However, the slit-width of $0.7''$ is
comparable to the typical effective radius of most of the galaxies
(\rad$\sim0.35''$; Fig.~\ref{fig:rad_smass}).  Most importantly, we
flux calibrate our spectroscopic measurements using total galaxy
fluxes anchored in deep ground and space-based photometry and confirm
that the uncertainty in the spectrophotometric calibration is $0.08$
mag \citep[see \S2.7 in][]{nanayakkara:16}.

\subsubsection{Galaxy Size at Fixed Stellar Mass}\label{sec:drad}

\begin{figure}
\figurenum{8}
\plotone{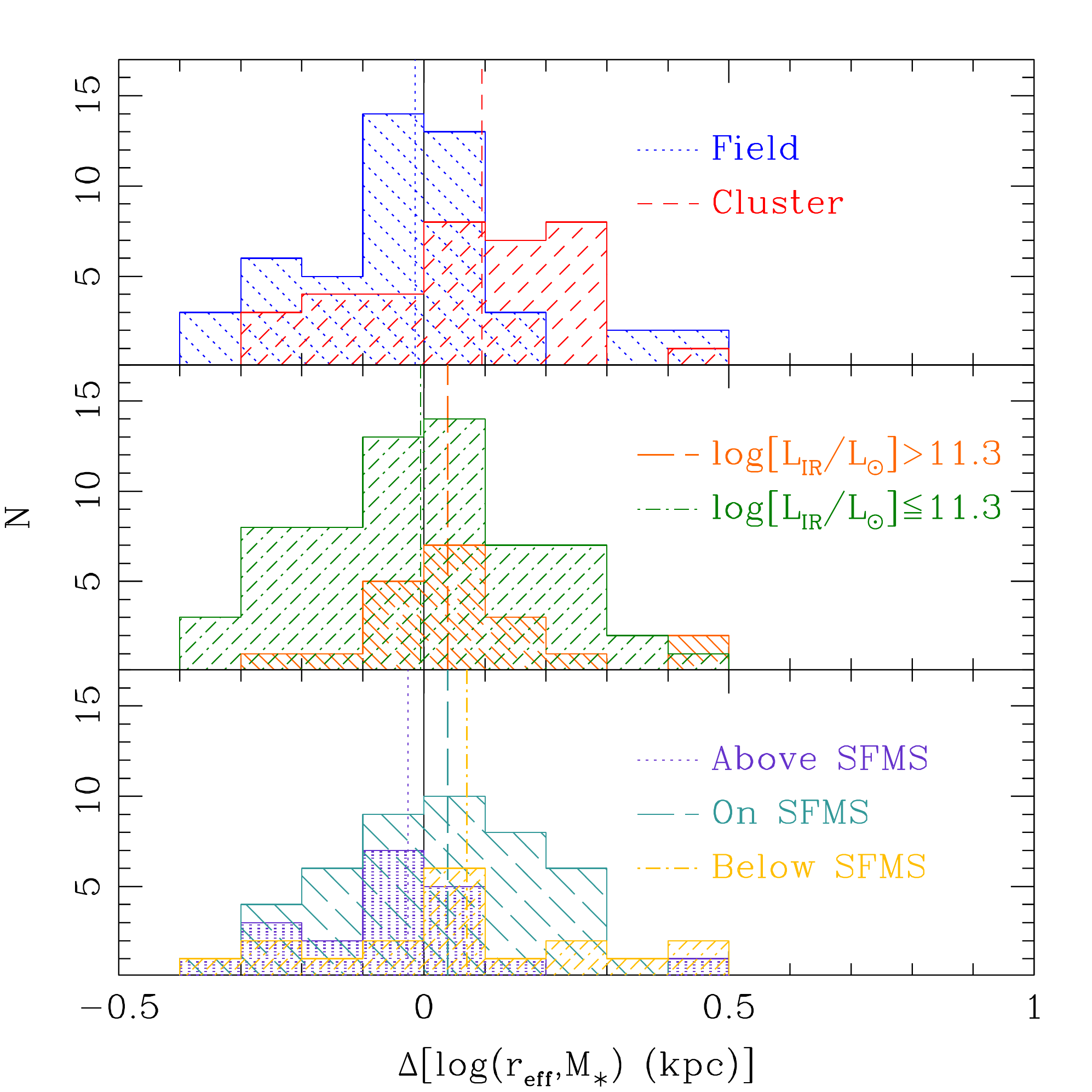}
\caption{The same as Fig.~\ref{fig:hist_rad} but showing the
  difference in \rad\ at a fixed stellar mass.  Here \drad\ is
  determined using the ($2\sigma$ clipped) least squares fit to
  \rad-\Mstar\ of our combined cluster and field galaxies
  (Fig.~\ref{fig:rad_smass}, black line in right panel).  Compared to
  their absolute \rad\ distributions (Fig.~\ref{fig:hist_rad}), K-S
  tests now measure higher likelihoods that the cluster and field
  galaxies are drawn from different \drad\ parent populations
  ($p=0.01$); this is also true for the above vs. below SFMS galaxies
  ($p=0.05$).  The \drad\ distributions of  
  the low-IR and LIRGs are more similar ($p=0.06$).
\label{fig:hist_drad}}
\end{figure}

To identify more subtle differences in galaxy size at fixed stellar
mass, we first fit a ($2\sigma$ clipped) least squares to
\rad-\Mstar\ using our combined cluster and field sample:

\begin{equation}
\Delta [\log ({\rm r}_{\rm eff},{\rm M}_{\star})] 
= \log ({\rm r}_{\rm eff},{\rm M}_{\star}) -
[ (0.253\times {\rm M}_{\star}) - 2.12]
\end{equation}

Our fitted least squares is virtually the same as the relation
measured by \citet{vanderwel:14} for galaxies at $z=2.0$
(Fig.~\ref{fig:rad_smass}, right).

When controlling for stellar mass, we find that the
\drad\ distributions for the cluster and field galaxies are likely
drawn from different parent populations (Fig.~\ref{fig:hist_drad},
top; $p=0.01$); this is in contrast to no difference in their {\it
  absolute} \rad\ distributions (Fig.~\ref{fig:hist_rad}).  At fixed
\Mstar, \Halpha-emitting cluster galaxies are $\sim0.1$~dex larger
than their field counterparts.  Our result is consistent with
\citet{allen:15} who find that star-forming cluster galaxies as
identified by their $UVJ$ colors are $\sim12$\% larger than those in
the field.

There is also a higher likelihood that at fixed stellar mass, galaxies
above the SFMS are drawn from a different \drad\ parent population
than those below (Fig.~\ref{fig:hist_drad}, bottom; $p=0.05$).  The
+SFMS galaxies are $\sim0.1$~dex smaller at a fixed \Mstar\ compared
to --SFMS galaxies (Fig.~\ref{fig:rad_smass}).  The compact nature of
the +SFMS galaxies across the entire stellar mass range suggests that
their star formation is more centralized than in the --SFMS galaxies
(see also \S\ref{sec:tracking}).

A K-S test of the \drad\ distributions for the low-LIR vs. LIRGs
measures $p=0.06$ which is not as statistically significant as when
comparing their absolute \rad\ distributions ($p=9.6\times10^{-6}$).
Because LIRGs are more massive (Fig.~\ref{fig:sfr_smass}), they also
tend to have larger radii.  Thus controlling for stellar mass reduces
differences in the LIRG and low-LIR populations.

\subsection{Galaxy Morphology \& Stellar Ages}

Having measured S\'ersic indices for 83 galaxies in our
\Halpha-emitting sample, we can compare the galaxy morphologies of the
different populations.  We find that all the galaxy populations (field
vs. cluster, LIRG vs. low-IR, above/on/below SFMS) have comparable
distributions in S\'ersic index as measured by a K-S test.  Most of
the galaxies (71/83) are disk-dominated systems ($n\leq2$).

The SED-based ages from \zfourge\ \citep{straatman:16} confirm that
the cluster and field galaxies have similar age distributions of
$\sim8.5$~Gyr.  This is also true for the LIRG and low-IR populations
(both are $\sim8.5$~Gyr).  However, comparison of the galaxies above
(+SFMS), on (=SFMS), and below (--SFMS) galaxies shows that their
average stellar ages increases from $\sim8.3$, $\sim8.6$, to
$\sim8.7$~Gyr respectively.  The younger light-weighted stellar
ages of the +SFMS galaxies is consistent with a starburst nature.

\subsection{Spatial Extent of \Halphased-Star Formation}

\begin{figure}
\figurenum{9}
\plotone{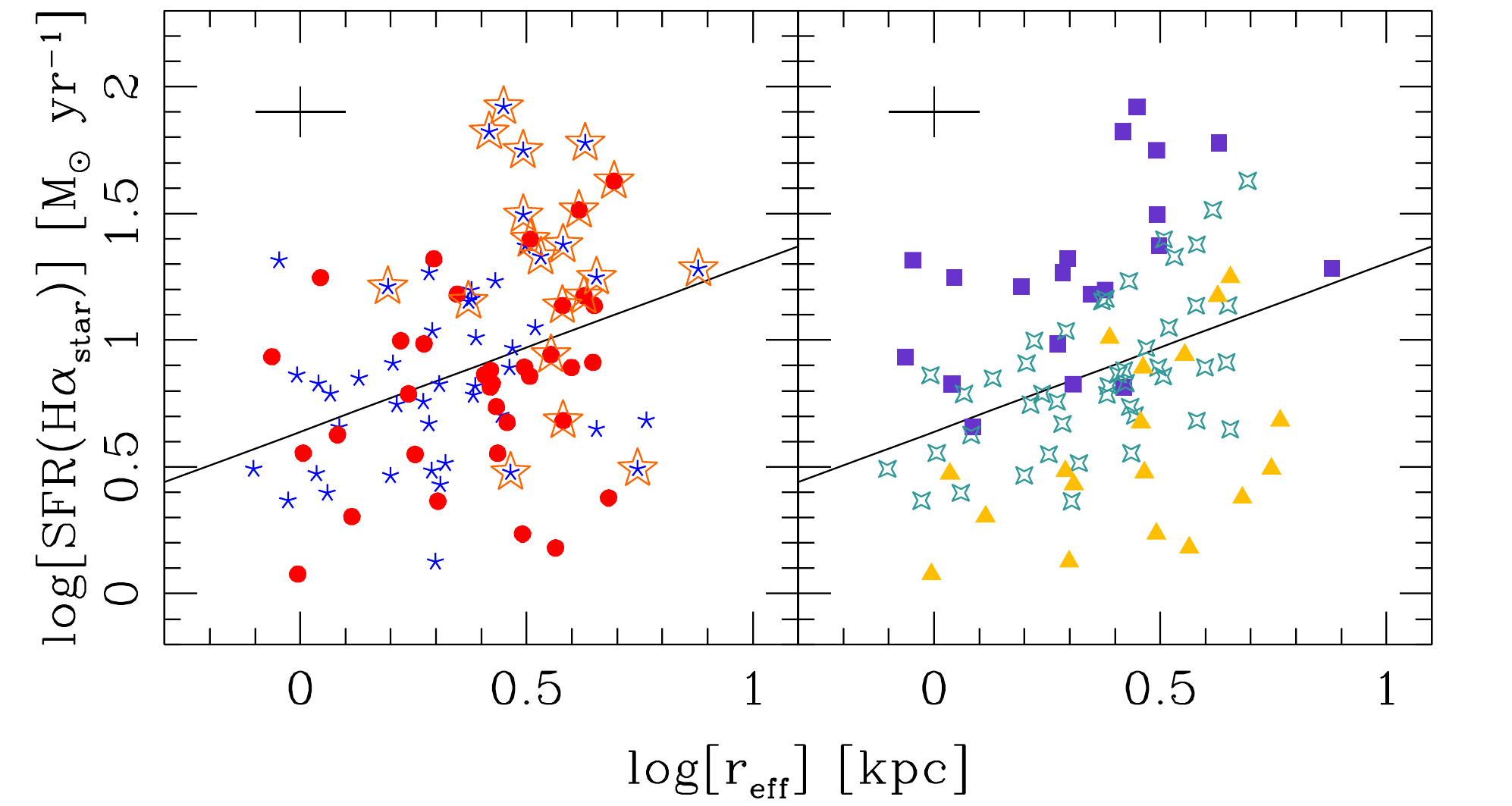}
\caption{{\it Left:} There are no differences in the cluster (filled
  red circles) and field (filled blue stars) galaxies when comparing
  their \Halphased\ star formation rate to their WFC3/F160W galaxy
  size.  The solid line in both panels is the least-squares fit
  ($2\sigma$ outliers removed) to the combined sample.  LIRGs (open
  orange stars) tend to be larger than low-IR galaxies in both
  environments.  {\it Right:} Galaxies above (filled squares), on
  (open crosses), and below (filled triangles) the \Halpha\ SFMS (see
  Fig.~\ref{fig:sfr_smass}, right) populate different regions: +SFMS
  galaxies have higher \Halphased-SFRs at a given size compared to
  --SFMS galaxies.
\label{fig:sfr_rad}}
\end{figure}

Using the star formation rates derived from \Halphased, the effective
radii measured using WFC3/F160W imaging, and stellar masses from SED
fitting, we first compare the \Halphased-SFR to galaxy size (\rad,
Fig.~\ref{fig:sfr_rad}; see \S\ref{sec:galfit} and Table~1).  Our
assumption that the \Halpha\ radii are comparable to the rest-frame
optical radii is supported by results from \citet[SINS;][]{forster:11}
who combined IFU and HST observations of six \Halpha-emitting
galaxies at $z\sim2$ and find no significant differences in their
sizes and structural parameters at these wavelengths.

The cluster and field galaxies have similar distributions, and
least-squares fits ($2\sigma$ outliers removed) confirm that both
populations have the same slopes within the errors.  As seen in
Fig.~\ref{fig:hist_rad}, the LIRGs tend to have larger \rad\ than the
low-IR galaxies because the LIRGs are more massive.  In
contrast, galaxies above the SFMS have higher \Halphased-SFRs at a
given size compared to those below the SFMS (Fig.~\ref{fig:sfr_rad}).

\begin{figure}
\figurenum{10}
\plotone{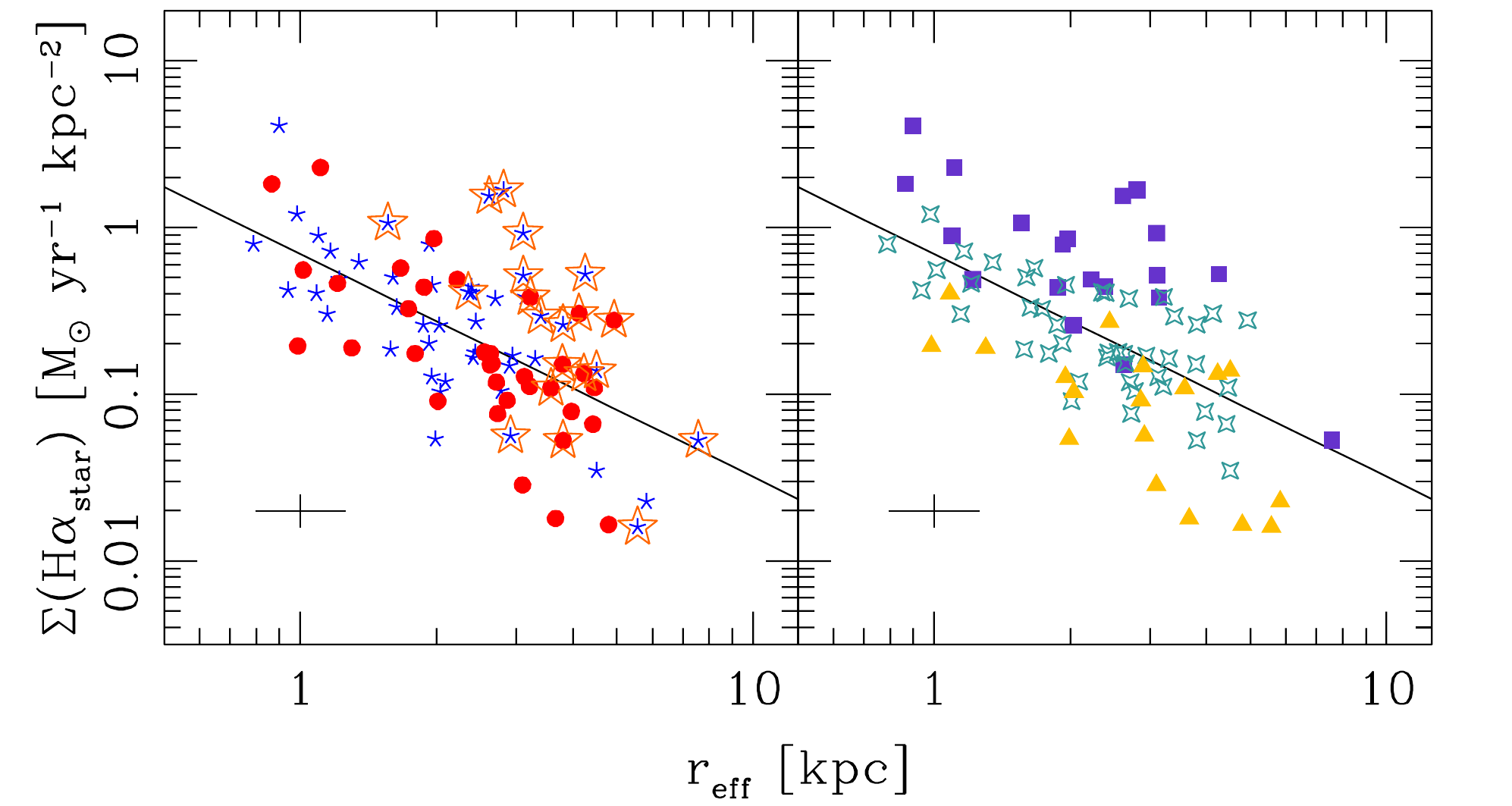}
\caption{{\it Left:} The star formation rate surface density
  \sigmaHa\ is measured with \Halphased\ star formation rate and
  WFC3/F160W galaxy size, and the solid line is the least-squares fit
  ($2\sigma$ outliers removed) to the combined sample.  Cluster
  (filled circles) and field (line stars) galaxies have the same
  distribution in \sigmaHa--\rad.  In contrast, the LIRGs (open stars)
  tend to be larger and have higher \sigmaHa\ compared to low-IR
  galaxies, $i.e.$ massive star-forming galaxies tend to have larger
  \rad\ and also be LIRGs.  {\it Right:} Galaxies above (filled
  squares), on (open crosses), and below (filled triangles) the
  \Halpha\ SFMS (see Fig.~\ref{fig:sfr_smass}, right) populate
  different regions: +SFMS galaxies are forming stars more intensely
  than --SFMS galaxies across the range in galaxy size.
\label{fig:sigsfr_rad}}
\end{figure}

\begin{figure}
\figurenum{11}
\plotone{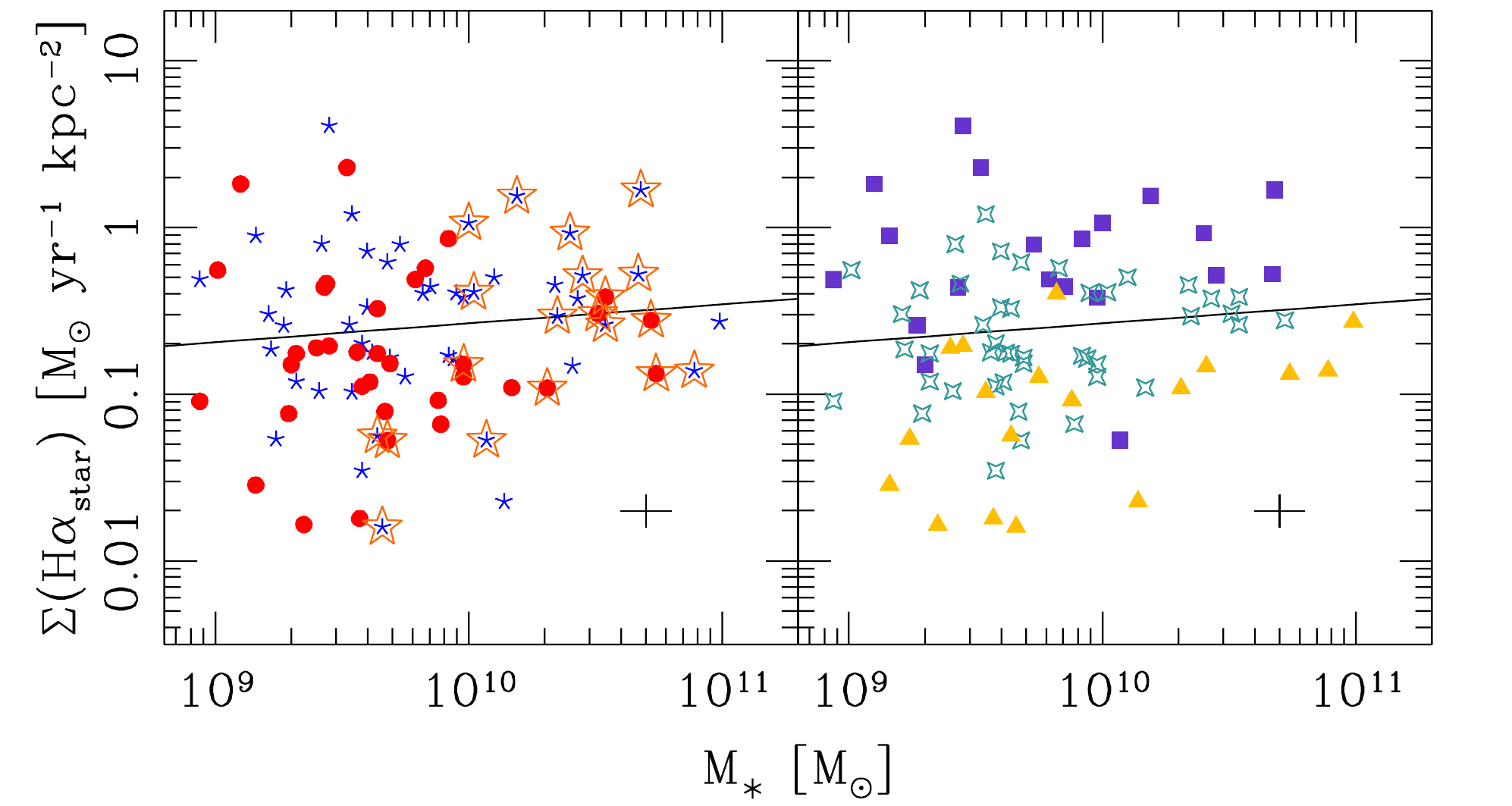}
\caption{{\it Left:} The star formation rate surface density
  \sigmaHa\ compared to stellar mass \Mstar\ where the solid line is
  the least-squares fit ($2\sigma$ outliers removed) to the combined
  sample.  Cluster (filled circles) and field (line stars) galaxies
  have the same distribution in \sigmaHa--\Mstar.  LIRGs (open stars)
  are more massive than low-IR galaxies, but both populations span the
  range in \sigmaHa.  {\it Right:} Galaxies above (filled squares), on
  (open crosses), and below (filled triangles) the \Halpha\ SFMS (see
  Fig.~\ref{fig:sfr_smass}, right) populate different regions: +SFMS
  galaxies are forming stars more intensely than --SFMS galaxies
  across the range in stellar mass.
\label{fig:sigsfr_smass}}
\end{figure}

We find similar results when comparing the star formation rate surface
density [\sigmaHa; see Eq.~4] to galaxy size (\rad,
Fig.~\ref{fig:sigsfr_rad}) and stellar mass (\Mstar,
Fig.~\ref{fig:sigsfr_smass}).  The cluster and field galaxies have
similar distributions, and least-squares fits ($2\sigma$-clipped) to
\sigmaHa-\rad\ and \sigmaHa-\Mstar\ confirm that both populations have
the same slopes within the errors.  Note that our sample spans a range
in galaxy size [$0.5<$\rad~(kpc)$<8$], star formation rate surface
density [$0.01<$\sigmaHa$<5$] where the units are \Msunyr~kpc$^{-2}$,
and stellar mass [$9<$\logMstarMsun$<11$].

In contrast, the LIRGs and low-IR populations are different: at a
given galaxy size, LIRGs tend to have higher SFR surface densities
(Fig.~\ref{fig:sigsfr_rad}, left).  As noted in \S3.3.2, LIRGs also
are typically $\sim5$ times more massive (Fig.~\ref{fig:sigsfr_smass})
and physically larger by $\sim70$\%. However, LIRGs are not all
starbursts, $i.e.$ LIRGs are found above, on, and below the SFMS
(Fig.~\ref{fig:sfr_smass}).

If we consider instead galaxies that lie above the SFMS, these +SFMS
systems have higher SFR surface densities than --SFMS galaxies
(Fig.~\ref{fig:sigsfr_rad}, right).  {\it At a given stellar mass},
the +SFMS galaxies tend to have smaller radii
(Fig.~\ref{fig:rad_smass}) and higher
\sigmaHa\ (Fig.~\ref{fig:sigsfr_smass}) compared to galaxies on/below
the SFMS.  Our results suggest that the \Halpha\ star formation in
+SFMS is more concentrated than those on/below the SFMS.

\subsection{\rhapsody\ Simulations: SFR-\Mstar}

We compare our measured \Halphased\ star formation rate versus stellar
mass relation to predictions from the \rhapsody\ simulations of
massive galaxy clusters \citep[M$_{\rm vir}>6\times10^{14}$~\Msun\ at
  $z=0$;][] {hahn:15,martizzi:16}.  These cosmological hydrodynamical
zoom-in simulations (R4K resolution) use the \ramses\ adaptive mesh
refined (AMR) code \citep{teyssier:02} to reach a spatial resolution
of 3.8\hinverse~kpc (physical), dark matter particle mass resolution
of $8.22\times10^8$\hinverse~\Msun, and baryonic mass resolution of
$1.8\times10^8$\hinverse~\Msun.  The simulations assume the standard
$\Lambda$CDM cosmology ($\Omega_{\rm M}=0.25$,
$\Omega_{\Lambda}=0.75$, $\Omega_{\rm b}=0.045$, $h=0.7$) and include
gas cooling, star formation, metal enrichment, and feedback from
supernovae and AGN.

The \rhapsody\ cluster simulations are well-matched to our COSMOS
cluster at \zcl.  As detailed in \citet[see their \S4]{yuan:14}, its
measured velocity dispersion of $\sigma_{1D}=552$~\kms\ corresponds to
a virial mass of $\log($M$_{\rm vir}/$M$_{\star})\sim13.5$.  Merger
trees from the GiggleZ Gpc simulation \citep{poole:15} shows that such
systems grow into a Virgo-mass cluster with
$\log($M$_{\rm vir}/$M$_{\star})\sim14.4$ by $z\sim0$.

We consider only simulated cluster galaxies at $z=2$ with star
formation rates $>1$~\Msunyr; these galaxies have stellar masses of
\logMstarMsun$=9-12$.  Here we assume that selecting by SFR is
equivalent to the instantaneous observed SFR as measured by
\Halphased.  We cannot apply the same observed $UVJ$ selection as
rest-frame colors are not available for the simulated galaxies.

From three \rhapsody\ cluster realizations, the least-squares fit to
the SFR-\Mstar\ relation is:

\begin{equation}
\log[ {\rm SFR(M}_{\odot}~{\rm yr}^{-1}) ] = 1.08 \left\{ \log [{\rm
M}_{\star}({\rm M}_{\odot})] -10\right\}  + \log(4.5)
\end{equation}

%\log [{\rm M}_{\star}({\rm M}_{\odot}) ] -10 \right} + \log(4.5)

% SFR = 4.5 \pm 0.6 M_sun/yr * (M*/1e10 M_sun)^(1.08 \pm 0.07)

The \rhapsody\ slope to the SFR-\Mstar\ relation is steeper than that
of the observed galaxies at $z\sim2$: 1.08 vs. 0.61
(Fig.~\ref{fig:sfr_smass}, right panel: gray and cyan lines
respectively).  Although the slopes are consistent within the scatter
of the simulations and observations (see \S\ref{sec:sfr_smass}), the
SFRs predicted by \rhapsody\ are lower by a factor of $\sim2$ for most
of the observed galaxies.  This difference between predicted and
observed SFRs at a given stellar mass ($i.e.$ the specific SFR) is
known to exist for field comparisons \citep[$e.g.$][]{dave:16b}. Here
we show that this discrepancy extends to the cluster environment as
well, $i.e.$ simulations over-predict how efficiently galaxies quench
at a given stellar mass for both the cluster and field environments
\citep[see][and references therein]{somerville:15}.  In the case of
\rhapsody, star formation histories at high redshift are slightly
under-resolved due to the mass resolution.  Future simulations with
higher resolution combined with multi-epoch observations are needed to
improve galaxy formation modeling at $z\sim2$.  We will explore more
key scaling relations compared to simulations in future work.

%------------------------------------------------------------
\section{Discussion}

Our analysis focuses on comparing at $z\sim2$ (i) cluster to field
galaxies (\ncl\ vs. \nfd); (ii) galaxies with \lir$>$\lircut (LIRGs)
to the low-IR population (\nirbright\ vs. \nirfaint); and (iii)
galaxies above, on, and below the \Halpha\ star-forming main sequence
(SFMS).  Because ours is a stellar mass-selected sample of
\Halpha-emitting galaxies [\logMstarMsun$>9$; see
  Fig.~\ref{fig:sfr_smass}], we are not limited to the high mass end
of the galaxy population.  We consider only \Halpha-selected galaxies
at $1.9<z<2.4$ because the redshifts for the quiescent galaxies are
based on photometry and/or grism spectroscopy
\citep{tomczak:14,momcheva:16}, neither of which are as precise as our
Keck/MOSFIRE redshifts determined with \Halpha.  We confirm that
selecting field galaxies using a more stringent cut of $>8\sigma_{\rm
  1D}$ from the mean cluster redshift does not change our results.

\subsection{\Halpha-Emitting Galaxies: Little Evidence of Environmental 
Dependence at $z\sim2$}

Our original motivation was to quantify how galaxy properties vary
with environment at $z\sim2$.  However, we find little evidence for
environmental dependence in \Halpha-emitting galaxies at $z\sim2$.  We
consistently measure the same relations for cluster and field galaxies
when comparing their \Halphased\ star formation rate to stellar mass
(Fig.~\ref{fig:sfr_smass}), galaxy size to stellar mass
(Figs.~\ref{fig:rad_smass} \& \ref{fig:hist_rad}), and star formation
concentration (Fig.~\ref{fig:sfr_rad}, \ref{fig:sigsfr_rad}, \&
\ref{fig:sigsfr_smass}).  The fraction of LIRGs and their median \lir\
are also the same in the cluster and field (\S\ref{sec:IRfrac}).  In
our study, the only measureable difference is that at fixed stellar
mass, \Halpha-emitting cluster galaxies are $\sim0.1$~dex larger than
the field (Fig.~\ref{fig:hist_drad}).

In terms of their physical properties, the \Halpha-emitting cluster
galaxies at \zcl\ are essentially the same population as the field.
This is consistent with our results in \citet{kacprzak:15} showing
that these very same cluster and field galaxies also follow the same
gas-phase metallicity versus stellar mass relation (MZR).  In
addition, we find no evidence for an environmental dependence when
comparing their kinematic scaling relations
\citep{alcorn:16,straatman:16}.

The handful of existing studies on galaxy overdensities at $z\gtrsim2$
similarly find little evidence for environmental effects.  Using
narrow-band imaging, \citet{koyama:13a} measure the same
SFR-\Mstar\ relation for \Halpha-emitters in a $z=2.16$ proto-cluster
as in the field.  Using high resolution imaging from the \hubble,
\citet{peter:07} measure the same size (radius) distributions for
field and proto-cluster galaxies at $z=2.3$.

In contrast, \citet{papovich:12} find that quiescent cluster galaxies
at $z=1.62$ are larger than their field counterparts, and
\citet{quadri:12} find a higher fraction of quiescent galaxies in the
same cluster.  Several studies also find evidence of enhanced star
formation in cluster galaxies at $z<2$
\citep{tran:10,brodwin:13,santos:14,tran:15,webb:15}.  The lack of
convincing evidence for strong environmental effects at $z\gtrsim2$
combined with the increasing differences between cluster and field
galaxies at lower redshifts pinpoints to $1.5\lesssim z\lesssim2$ as
the critical epoch for ending star formation in cluster galaxies and
building the spheroid population in clusters.

\subsection{Tracking Galaxy Growth with the \Halpha\ Star-Forming Main
  Sequence} \label{sec:tracking}

Given the physical properties of \Halpha-emitting galaxies show little
environmental dependence (see above), we can use the combined cluster
and field sample at $z\sim2$ to compare galaxies above, on, and below
the star-forming main sequence (SFMS) as well as compare the
IR-luminous (\nirbright; LIRG) to low-IR (\nirfaint) populations
(Figs.~\ref{fig:sfr_smass} through \ref{fig:sigsfr_smass}). Because
our spectroscopic target selection is based on \zfourge, we are
mass-limited to \logMstarMsun$\sim9$ at $z\sim2$
\citep{tomczak:14,nanayakkara:16}.

Being a LIRG does not necessarily mean the galaxy is a starburst
because LIRGs are found above, on, and below the SFMS
(Fig.~\ref{fig:sfr_smass}, left).  Rather, IR-luminosity tends to
track closely with stellar mass such that massive galaxies
[\logMstarMsun$>10$] tend to be LIRGs.  On average, LIRGs are $\sim5$
times more massive and $\sim70$\% larger than low-IR galaxies
(Figs.~\ref{fig:rad_smass}, \ref{fig:hist_rad}, \& \ref{fig:sfr_rad}).
When controlling for stellar mass, there is less difference in the
size distributions the LIRGs and low-IR galaxies
(Fig.~\ref{fig:hist_drad}).  Note that the mass range of our
\Halpha-emitting galaxies reaches \logMstarMsun$\sim9$, $i.e.$ a
factor of about $5-10$ times lower than previous studies that compared
LIRGs to the general galaxy population at $z>1$
\citep[$e.g.$][]{swinbank:10}.

In terms of tracking how galaxies grow, systems that lie above the
\Halpha\ star-forming main sequence (+SFMS) have smaller radii at a
given stellar mass than those that are below the SFMS
(Fig.~\ref{fig:rad_smass}, right; see \S\ref{sec:sfms}).  The +SFMS
galaxies tend to have higher \Halphased-SFRs at a given galaxy size
(Fig.~\ref{fig:sfr_rad}) and higher \Halphased\ SFR surface densities
compared to those below [\sigmaHa; Figs.~\ref{fig:sigsfr_rad} \&
\ref{fig:sigsfr_smass}], $i.e.$ their star formation is more compact.
The +SFMS galaxies also have younger SED-based stellar ages of
$\sim8.3$~Gyr compared to $\sim8.7$~Gyr for --SFMS galaxies.  Taken as
a whole, our results indicate that +SFMS are starbursts with
\Halpha\ star formation concentrated in their cores \citep[see
  also][]{barro:15}.

At $z\sim1$, field galaxies are preferentially growing their disks
\citep{nelson:16a}.  In combination with our observations indicating
that starbursts at $z\sim2$ are growing their stellar cores, these
results suggest a sequence where +SFMS galaxies are building up
their stellar cores at $z\sim2$ and then their stellar disks at
$z\sim1$, $i.e.$ inside-out growth, likely by continuing gas accretion
at $z<2$ \citep[$e.g.$][]{kacprzak:16}.  Such a scenario naturally
produces the older stellar populations of bulges relative to disk.
This can also explain the rise of spheroids in clusters if the cluster
environment prevents the growth of stellar disks even as star
formation in the galaxies' cores is quenched at $z<1.5$
\citep{brodwin:13,tran:15}.  While our hypothesis is based on the
  +SFMS galaxies, we note that galaxies at $z\sim2$ in general must
  grow physically larger by $z\sim1$ \citep[$e.g.$][]{vanderwel:14}.

\subsection{Star Formation Rates at $z\sim2$: Caveat Emptor}

Our analysis is based on the relative comparison of cluster and field
galaxies where properties for both are determined in the same manner.
Thus our results do not depend on the absolute conversion of, $e.g.$
\Halpha\ flux to SFR.  However, we do find that the \Halphased\ SFRs
are offset from \lir\ SFRs (Fig.~\ref{fig:Ha_lir}).  The large
uncertainty and likely offset from relations measured at $z\sim0$
brings into question our ability to measure reliable SFRs at $z>1$.

There are several ongoing efforts to better understand star formation
and dust laws at $z>1$ that should help with calibrating existing
relations.  Recent studies at $z\sim2$ find evidence of changing
ionization conditions \citep{sanders:16a} as well as different dust
laws \citep{reddy:15,forrest:16,shivaei:16} that can be incorporated
into models.  However, until we identify a more robust method for
measuring SFRs in the distant universe, direct comparisons between
studies will require carefully accounting for different methods of
measuring SFRs.

%------------------------------------------------------------
\section{Conclusions}

Our \zfire\ program combines Keck/MOSFIRE spectroscopy with the wealth
of multi-wavelength observations available in the COSMOS legacy field
to explore galaxy scaling relations as a function of environment at
$z\sim2$.  Our advantage is that we select galaxies at $z\sim2$ based
on their stellar masses as measured by \zfourge, a deep imaging survey
that uses medium-band NIR filters to obtain high-precision photometric
redshifts \citep[$\sigma_z\sim0.02$;][]{straatman:16}. We focus on the
spectroscopically confirmed galaxy cluster at $z=2.095$ in the COSMOS
legacy field \citep{spitler:12,yuan:14} and compare to the field
population at $z\sim2$.

In comparing \Halpha-emitting cluster (\ncl) and field (\nfd) galaxies
[\logMstarMsun$>9$; AGN removed], we find little evidence of
environmental influence on any of the galaxy scaling relations.  Both
cluster and field populations are consistent with published relations
between star formation rate and stellar mass (SFR-\Mstar;
Fig.~\ref{fig:sfr_smass}) as well as galaxy size and stellar mass
(\rad-\Mstar; Fig.~\ref{fig:rad_smass}) at $z\sim2$.  The cluster and
field populations also have the same distribution when comparing their
\Halphased-SFR surface density [\sigmaHa] to galaxy size and stellar
mass (Figs.~\ref{fig:sigsfr_rad} \& \ref{fig:sigsfr_smass}).  The
results in this analysis mirror our existing \zfire\ results that show
these same cluster and field galaxies have the same gas-phase
metallicity vs. stellar mass relation \citep{kacprzak:15}, kinematic
mass vs. stellar mass \citep{alcorn:16}, and ISM conditions
\citep{kewley:16}.   The only subtle indication of possible
  environmental dependence is that at fixed stellar mass, the 
  \Halpha-emitting cluster galaxies are $\sim0.1$~dex larger than in 
  the field (Fig.~\ref{fig:hist_drad}).

Using \spitzer/\mipsmu\ observations, we identify \nirbright\ galaxies
with \lir$>$\lircut, $i.e.$ Luminous Infra-Red Galaxies (LIRGs).  Note
that our mass range of \logMstarMsun$\sim9$ is a factor of about
$5-10$ times lower than previous studies that compared LIRGs to the
general galaxy population at $z>1$ \citep[$e.g.$][]{swinbank:10}.
The LIRG fraction is comparable within errors between the cluster
  and the field (19\% and 26\% respectively), and we do not find any
indication that LIRGs in the cluster are different from those in the
field.  IR-luminosity tracks with stellar mass such that our most
massive galaxies [\logMstarMsun$>10$] are dominated by LIRGs.  As a
result, LIRGs tend to be $\sim5$ times more massive with radii that
are $\sim70$\% larger than low-IR galaxies (\rad$\sim3.8$~kpc
vs. $\sim2.0$~kpc; Fig.~\ref{fig:hist_rad}).  The LIRGs are not all
starbursts as they are found above, on, and below the
\Halpha\ Star-Forming Main Sequence (SFMS; Fig.~\ref{fig:sfr_smass}).

We show that separating galaxies into those above the
\Halpha\ Star-Forming Main Sequence (+SFMS), on, and below (--SFMS)
provides insight into how galaxies grow (Fig.~\ref{fig:sfr_smass}).
Galaxies in the three groups span the full range in parameter space,
but the +SFMS galaxies have smaller radii at a given stellar mass
compared to --SFMS (Fig.~\ref{fig:rad_smass}).  The +SFMS galaxies
also tend to have higher SFR surface densities compared to galaxies
with depressed SFRs (Figs.~\ref{fig:sfr_rad}, \ref{fig:sigsfr_rad}, \&
\ref{fig:sigsfr_smass}), and younger SED-based stellar ages compared
to galaxies below the SFMS ($\sim8.3$~Gyr vs. $\sim8.7$~Gyr).

These lines of evidence indicate that +SFMS galaxies (starbursts) have
concentrated \Halpha\ star formation and are actively growing their
cores at $z\sim2$.  We infer that while starbursts in the field go on
to grow their stellar disks at $z\sim1$ \citep{nelson:16a}, cluster
starbursts are likely to be quenching their star formation at $z<2$
\citep{brodwin:13,tran:15} to then evolve into quiescent spheroids
\citep{quadri:12,papovich:12}.

We compare the \Halpha\ SFR-\Mstar\ relation to predictions from the
\rhapsody\ simulations of massive galaxy clusters
($>6\times10^{14}$~\Msun\ at $z=0$) based on the \ramses\ Adaptive
Mesh Refinement code.  We find that the predicted slope for the
SFR-\Mstar\ relation is steeper than the observed values (1.08
vs. 0.61), and that the predicted SFRs are $\sim2$ times lower than
observed.  Simulations in general continue to over-predict how
efficiently galaxies quench at a given stellar mass in both the
cluster and field environments.  We will continue to explore how
observed galaxy scaling relations compare to simulations in future
work.

On a cautionary note, there is considerable scatter and likely offset
in star formation rates based on \Halpha\ and those based on IR
luminosity (or UV+IR) at $z\sim2$ (Fig.~\ref{fig:Ha_lir}).  This is in
contrast to the relatively small scatter at $z\sim0$ between
\Halpha\ and \mipsmu\ derived star formation rates for activity at
$<100$~\Msunyr\ \citep[$e.g.$][]{hao:11}.  It is sobering to consider
the large uncertainty in measuring robust SFRs, especially at
higher redshifts when SFRs are increasing in general
\citep{garn:10,whitaker:14,tomczak:16}.  Our ability to accurately
measure star formation at $z\gtrsim2$ is likely to be limited due to,
$e.g.$ our understanding of how ionization conditions evolve.  However,
we stress that the strength of this analysis lies in using the same
observables to directly compare across different galaxy populations at
$z\sim2$.

In a companion \zfire\ paper, we estimate gas masses and gas depletion
timescales for the same cluster and field galaxies at $z\sim2$.
Ongoing analyses also include a comparison of the Tully-Fisher
relation (Straatman et al., submitted) and constraints on the Initial
Mass Function (Nanayakkara et al., in prep).  By measuring galaxy
scaling relations for cluster and field galaxies at $z\sim2$, \zfire\
provides a unique benchmark for quantifying galaxy evolution as a
function of environment.

%------------------------------------------------------------

\include{table1-short}

%\include{table1-galaxies}

%------------------------------------------------------------
\acknowledgements

We are grateful to the MOSFIRE team with special thanks to M. Kassis,
J. Lyke, G. Wirth, and L. Rizzi on the Keck support staff.  K. Tran
thanks M. Kriek, A. Shapley, S. Price, B. Forrest, and Jimmy for
helpful discussions.  We also thank the referee for a thoughtful and
constructive report.  This work was supported by a NASA Keck PI Data
Award administered by the NASA Exoplanet Science Institute. Data
presented herein were obtained at the W. M. Keck Observatory from
telescope time allocated to NASA through the agency's scientific
partnership with the California Institute of Technology and the
University of California. The Observatory was made possible by the
generous financial support of the W. M. Keck Foundation.  K. Tran
acknowledges that this material is based upon work supported by the
National Science Foundation under Grant Number 1410728.  GGK
acknowledges the support of the Australian Research Council through
the award of a Future Fellowship (FT140100933).  The authors wish to
recognize and acknowledge the very significant cultural role and
reverence that the summit of Mauna Kea has always had within the
indigenous Hawaiian community. We are most fortunate to have the
opportunity to conduct observations from this mountain.

\bibliographystyle{/Users/vy/Work/aastex/aastexv6.0/aasjournal}

\bibliography{/Users/vy/Work/aastex/tran}

\end{document}

%% file: table1-short.tex
\floattable
\begin{deluxetable}{rrrrrrrrrrrrrrr}
\rotate
\tabletypesize{\scriptsize}
\tablecaption{Galaxy Properties\tablenotemark{a}}\label{tab:galaxies}
\tablewidth{0pt}
\tablehead{
\colhead{\zfire\tablenotemark{b}}            &
\colhead{\zfourge\tablenotemark{b}}          &
\colhead{$\alpha(2000)$}        &
\colhead{$\delta(2000)$}        &
\colhead{\zspec}                &
\colhead{f\Halpha\tablenotemark{c}} &
\colhead{err(f\Halpha)\tablenotemark{c}} &
\colhead{$\log$(\lir/\Lsun)\tablenotemark{d}}          &
\colhead{\logMstarMsun}         &
\colhead{A$_{\rm V,star}$}      &
\colhead{$\log(t_{\rm star})$\tablenotemark{e}}  &
\colhead{SFR(\Halphased)\tablenotemark{f}} &
\colhead{S\'ersic $n$}          &
\colhead{\rad\ ($''$)}  &
\colhead{Pflag\tablenotemark{g}}
}
\startdata
%   _ID_v2       _id_v3          _RA         _DEC     _zspec       _fha       _fha_err	    _llir       _lmass_Avsed_v3    _lage     _sfrHa _nsersic_reff_arcsec   _morph_flag
       237 &        912 &  150.19057 &    2.18848 &   2.1572 &     1.46 &     0.17 & 	  \nodata &       9.65 &    0.6 &    8.1 &      6.0 &  \nodata &  \nodata &    -99 \\ 
       342 &       1108 &  150.19051 &    2.19065 &   2.1549 &     3.98 &     0.09 & 	     11.93 &      10.45 &    1.1 &    8.9 &     31.3 &    0.8 &    0.4 &      0 \\ 
      1085 &       2114 &  150.18338 &    2.20192 &   2.1882 &     2.53 &     0.07 & 	  \nodata &       9.60 &    0.1 &    8.4 &      5.6 &    1.3 &    0.2 &      0 \\ 
      1180 &       2168 &  150.12984 &    2.20287 &   2.0976 &     1.33 &     0.15 & 	  \nodata &       8.94 &    0.0 &    8.1 &      2.3 &    4.0 &    0.2 &      2 \\ 
      1349 &       2517 &  150.20306 &    2.20554 &   2.1888 &     1.04 &     0.06 & 	     11.23 &       9.82 &    0.3 &    8.3 &      3.0 &    4.0 &    0.1 &      1 \\ 
      1385 &       2510 &  150.12344 &    2.20565 &   2.0978 &     3.28 &     0.23 & 	  \nodata &       9.30 &    0.1 &    8.0 &      6.5 &    0.5 &    0.3 &      0 \\ 
      1617 &       2989 &  150.09697 &    2.20917 &   2.1732 &     1.32 &     0.14 & 	     11.22 &      10.14 &    0.5 &    8.7 &      4.8 &    0.9 &    0.7 &      2 \\ 
      1814 &       3175 &  150.16809 &    2.21129 &   2.1704 &     4.55 &     0.10 & 	  \nodata &       9.95 &    0.4 &    8.5 &     14.6 &    1.0 &    0.3 &      0 \\ 
      2007 &       3375 &  150.16566 &    2.21366 &   2.0086 &     1.17 &     0.14 & 	     11.17 &       9.42 &    0.4 &    8.5 &      3.1 &    0.9 &    0.1 &      0 \\ 
      2153 &       3669 &  150.16533 &    2.21584 &   2.0123 &     4.86 &     0.18 & 	     11.48 &      10.07 &    0.7 &    8.8 &     19.2 &    4.0 &    0.9 &      2 \\ 
      2522 &       4084 &  150.19379 &    2.22011 &   2.1511 &     1.10 &     0.11 & 	     11.48 &       9.64 &    0.3 &    8.1 &      3.0 &    0.5 &    0.4 &      0 \\ 
\enddata
\tablenotetext{a}{\it The complete version of this table is available
       in electronic format.}
\tablenotetext{b}{We list galaxy identification numbers
       from \zfire\ \citep{nanayakkara:16}
       and \zfourge\ \citep{straatman:16}.  We include only galaxies
       with spectroscopic redshift quality flag of
       $Q_z=3$ \citep{nanayakkara:16} and $1.9<$\zspec$<2.4$. Cluster
       members have $2.08<$\zspec$<2.12$ \citep{yuan:14}.}
\tablenotetext{c}{Observed \Halpha\ fluxes and errors are in units of
$10^{-17}$~erg~s$^{-1}$~cm$^{-2}$. }
\tablenotetext{d}{In our analysis of IR-luminous vs. low-IR systems,
we select IR-luminous galaxies using $\log$(\lir/\Lsun)$>11.3$.}
\tablenotetext{e}{Stellar age in units of Gyr and based on SED fitting
with FAST \citep{kriek:09a}.}
\tablenotetext{f}{\Halphased\ star formation rates in units of \Msunyr\
and based on dust-corrected \Halpha\ fluxes (Eq.~\ref{eq:AHa};
see \S\ref{sec:dust}).}
\tablenotetext{g}{Pflag denotes quality of profile fit used
to measure the S\'ersic index $n$ and the effective radius \rad.
Pflag values are -99 (not fit), 0 (good fit), 1 (fair fit), and 2
(questionable fit).}
\end{deluxetable}